\documentclass[a4paper,11pt]{article}
\usepackage[dvipdfmx]{graphicx}
\usepackage{jheppub} 
\usepackage{lineno}
\nolinenumbers
\usepackage{indentfirst}
\usepackage{comment}
\usepackage{mathtools}

\newcommand{\nn}{\nonumber \\}

\newcommand{\w}{|w|}
\newcommand{\dn}{\frac{\partial}{\partial n}}

\arxivnumber{} 

\title{Consistency between Bulk and Boundary Causalities in Asymptotically Anti-de Sitter Spacetimes}

\author[a]{Lei Fu,}
\affiliation[a]{Department of Mathematics, Nagoya University, Nagoya 464-8602, Japan} 
  
\author[a,b]{Keisuke Izumi}
 \affiliation[b]{Kobayashi-Maskawa Institute, Nagoya University, Nagoya 464-8602, Japan}

\author[a]{and Daisuke Yoshida}

\emailAdd{fu.lei.v2@s.mail.nagoya-u.ac.jp} 
\emailAdd{FL514859503fulei@hotmail.com} 
\emailAdd{izumi@math.nagoya-u.ac.jp} 
\emailAdd{dyoshida@math.nagoya-u.ac.jp}

\abstract{
We investigate the consistency between bulk and boundary causalities in static, spherically symmetric, asymptotically anti-de Sitter (AdS) spacetimes.
We derive a general formula that provides sufficient conditions 
for time advance, where bulk propagation arrives earlier than any boundary propagation. 
As an application, we show that in Reissner--Nordstr\"{o}m--anti de Sitter spacetime, no geodesic satisfies the sufficient conditions for time advance even in the super-extremal case.  Furthermore, we demonstrate that the Einstein--Euler--Heisenberg theory exhibits time advance 
when one or a linear combination of the coupling constants is positive and below an upper bound determined by the AdS length scale.
}

\begin{document}
\maketitle
\flushbottom
\section{Introduction}
The anti-de Sitter/conformal field theory (AdS/CFT) correspondence~\cite{Maldacena:1997re,Gubser_1998,Witten:1998qj,Witten:1998zw} suggests that causal processes connecting boundary points through the bulk are somehow realized in the viewpoint of the boundary theory. 
Thus, the causal process through the bulk must also be causal from the viewpoint of the boundary theory.
From this perspective, comparing the causality in the bulk and on the boundary is a significant topic.
As suggested in Refs.~\cite{Woolgar1994ar,Page:2002xn}, the positivity of energy for asymptotically AdS spacetimes implies that bulk causality is consistent with that in the boundary theory.
Gao and Wald~\cite{Gao:2000ga} further showed that, if the null convergence condition~\footnote{
As noted in Ref.~\cite{Gao:2000ga}, the null convergence condition can be weakened to the averaged version in the sense of Borde~\cite{Borde:1987qr}. 
}, the null generic condition, and the global hyperbolicity of the conformally completed spacetime are satisfied in the bulk theory,  
such a consistency is guaranteed.
Engelhardt and Fischetti~\cite{Engelhardt2016aoo} later generalized the results of Gao and Wald, deriving a condition weaker than the null convergence condition that still ensures this consistency.
Based on these studies, the consistency between bulk and boundary causalities is not guaranteed if certain conditions are not satisfied.
Our objective is to investigate the conditions under which the causal relationships between the bulk and the boundary become inconsistent, a situation which we refer to as time advance, by perturbative analysis.

The method of characteristics~\cite{courant1989methods} reveals the causal structure of a given theory.
For instance, in general relativity, the fastest propagation occurs at the speed of light, which is defined as null geodesics with respect to the spacetime metric.
However, when these theories are extended to include the derivative corrections in the effective field theory approach, superluminal propagations, which follow  spacelike curves with respect to the spacetime metric, possibly arise as studied, for example, in Refs.~\cite{Scharnhorst:1990sr, Barton:1989dq, Barton:1992pq, Latorre:1994cv, Dittrich:1998fy, DeLorenci:2000yh} for the flat spacetime and 
 Refs.~\cite{Drummond:1979pp, Daniels_1994,Shore:1995fz, Daniels:1995yw, Shore:2007um, Cho:1997vg, Izumi:2014loa, Reall:2014pwa,Allahyari:2019jqz, Cao:2021sty, Reall:2021voz, Davies:2021frz} for curved spacetimes.
In such cases, the causal structure must be analyzed by using the fastest propagation, which is often described by null geodesics of an effective metric. 
Since time advance refers to the situation in which a bulk propagation reaches a boundary point earlier than any boundary-constrained propagation, one can investigate time advance by comparing the fastest propagation in the bulk and on the boundary, using the effective metric.
In Refs.~\cite{Brigante_2008,Buchel_2009,Buchel_2010,Camanho_2010,Camanho_20101,Camanho_2011, Andrade_2017}, it was shown that the graviton propagation in Gauss--Bonnet and Lovelock gravity can exhibit time advance, which is related to inconsistencies in the boundary theory, such as violation of the viscosity bound. 
The requirement of forbidding time advance imposes constraints on the parameters of these gravity theories.
See also Ref.~\cite{Camanho:2014apa} for discussions of time advance in the small impact factor limit, in the context of effective field theory approaches to on-shell graviton scattering amplitude.

Following the direction developed in Refs.~\cite{Brigante_2008,Buchel_2009,Buchel_2010,Camanho_2010,Camanho_20101,Camanho_2011, Andrade_2017}, in this paper, we examine the bulk propagations in general static, spherically symmetric, asymptotically AdS metrics,
rather than restricting ourselves to specific effective theories such as Gauss--Bonnet or Lovelock gravity. 
By comparing the fastest bulk propagation with the boundary causality, we establish the sufficient conditions for time advance which can be applied perturvatively to a given effective metric.
Subsequently, we apply our time advance conditions to the Einstein--Maxwell theory and the Einstein--Euler--Heisenberg theory in the presence of a negative cosmological constant. 
For the Einstein--Maxwell case, specifically, the Reissner--Nordstr\"{o}m--Anti-de~Sitter (RNAdS) spacetime, we find that no geodesic satisfies the sufficient conditions for time advance,
while, in the Einstein--Euler--Heisenberg theory, time advance is shown to occur when the parameters lie within a certain range.
Supposing that the time advance is prohibited in the Einstein--Euler--Heisenberg theory, the parameters must lie outside the region that leads to time advance, thereby imposing constraints on this theory.

This paper is organized as follows. In Sec.~\ref{sec:bb causality}, we introduce boundary and bulk causalities in asymptotically AdS spacetime and provide the sufficient conditions for time advance. In Sec.~\ref{sec:ta condition}, we derive the general formulas for time advance conditions in a general static, spherically symmetric, asymptotically AdS spacetime. Sec.~\ref{sec:application} applies these results to the exact examples. 
In Sec.~\ref{subsec:EM}, we analyze the static, spherically symmetric solutions of the Einstein--Maxwell theory, namely the RNAdS solution. Then we turn to the static, spherically symmetric solution of the Einstein--Euler--Heisenberg theory in Sec.~\ref{subsec:EH}.
Finally, we present a summary and discussion in Sec.~\ref{sec:summary}. The detailed calculations are presented in the Appendix. Throughout the paper, the unit  $c=1$ is used. The notation $(a,b,c,...)$ denotes $(t,r)$, $(i,j,...)$ refers to the coordinates of the sphere $S^{D-2}$ where $D$ is the dimension of the whole spacetime, and $(\mu,\nu,...)$ indicates the components of the whole spacetime. 


\section{Boundary/bulk causality and time advance}
\label{sec:bb causality}

The main focus of this paper is to compare boundary causality and bulk causality in asymptotically AdS spacetimes. 
Given a bulk field theory, the causality associated with its field equations can be derived.
In many cases, the boundary of the causally connected region is described by null geodesics with respect to an {\it effective metric}
\footnote{
If the kinetic terms of the wave equations are Klein--Gordon type, such as Einstein--Maxwell equations, the effective metric is simply given by the spacetime metric. However, as discussed in Appendix \ref{app:EFT}, once the higher derivative corrections to the Einstein--Maxwell equations are included, the effective metric generally differs from the spacetime metric. 
We assume that the effective metric asymptotically approaches AdS near the boundary.
}.
Throughout this paper, we focus on such cases.  In particular, we focus on the case where the effective metric is a $D$-dimensional asymptotically AdS metric, which can be written as
\begin{align}
    ds^2_{\text{AAdS}} \approx -\left(
    1+\frac{r^2}{\ell^2}
    \right)dt^2+\frac{dr^2}{1+\frac{r^2}{\ell^2}}+r^2d\Omega_{D-2}^2.
    \label{metAAdS}
\end{align}
Here, $d\Omega_{D-2}^2$ represents the metric for the unit $D-2$ sphere, $\ell$ denotes the AdS radius, and $\approx$ implies that sub-leading terms in the $r \rightarrow \infty$ limit are ignored.

\begin{figure}[t]
\centering
\includegraphics[width=1.0\textwidth]{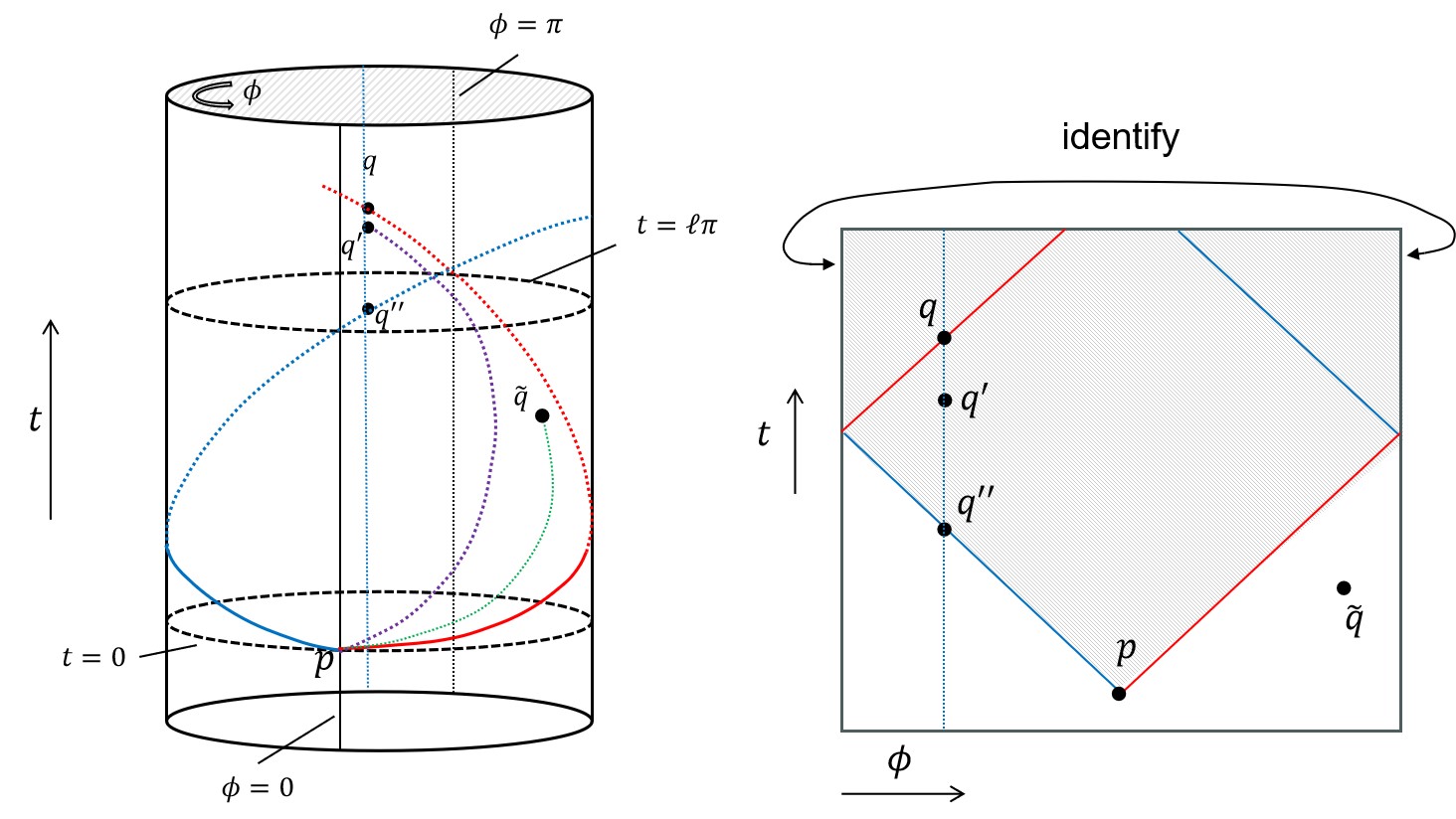}
\caption{The left figure depicts the asymptotically AdS spacetime, while the right figure presents an unfolded view of the left diagram. 
The red and blue curves represent null geodesics along the boundary, whereas the purple and green curves represent null geodesics within the bulk. 
In the right diagram, the gray region is causally related to $p$ from the perspective of boundary causality, while the white region is spacelike-separated. Consequently, the purple curve ends at $q'$ within the gray region, indicating that no time advance occurs. In contrast, the green curve terminates at $\tilde{q}$ in the white region, leading to a time advance.}
\label{AdS1}
\end{figure}

In the asymptotic region of Eq.~\eqref{metAAdS}, the foliation by the $r$ - constant $D-1$ dimensional hypersurfaces $\{\Sigma_{r}\}$ can be taken.
The induced metric on $\Sigma_{r}$, provided that $r$ is sufficiently large, is approximately given by
\begin{align}
    &ds^2_{\Sigma_{r}}\approx\frac{r^2}{\ell^2} ds^{2}_{\text{ESU}}, 
\end{align}
where $ds^2_{\text{ESU}}$ is the metric of the Einstein static universe
 given by
\begin{align}
      &ds^{2}_{\text{ESU}} \coloneqq - dt^2+ \ell^2d\Omega_{D-2}^2.
      \label{defESU}
\end{align}
The AdS boundary $\partial \mathcal{M}$ is defined by $r \rightarrow \infty$ limit of $\Sigma_{r}$.
Since a null geodesic with respect to the boundary metric $ds^2_{\partial \mathcal{M}}$ is also a null geodesic with respect to the conformally related metric $ds^{2}_{\text{ESU}}$, we will define the boundary causality by that with respect to the metric of the Einstein static universe \eqref{defESU}.
More precisely, we say that {\it $p, q \in \partial \mathcal{M}$ are causally connected in the sense of boundary causality} if there exists a causal curve that connects $p$ and $q$ and is contained in $\partial \mathcal{M}$.
In addition, we say that {\it the boundary causality is inconsistent with the bulk causality} if 
there is a causal curve in $\mathcal{M}$ that connects the points $p, q \in \partial \mathcal{M}$ which are not causally connected in the sense of boundary causality (see Fig.~\ref{AdS1}).  

Let us analyse the causal structure in more detail.
Let $\gamma(p,q)$ be a null geodesic from a point $p$ on the boundary $\partial \mathcal{M}$ to another point $q$ on the boundary $\partial \mathcal{M}$. 
Suppose the coordinates of the unit $D-2$ sphere $d\Omega_{D-2}^2$ in Eq.~\eqref{metAAdS} are written by the polar angles $\theta_{1}, \theta_{2}, \cdots, \theta_{D-3}$ and the azimuthal angle $\phi \in (- \pi, \pi]$.
Without loss of generality, $p$ and $q$ are on the plane defined by $\theta_{1} = \theta_2 = \dots = \theta_{D-3} = \pi/2$. 
Moreover, we can set $t = \phi = 0$ at the point $p$ and define $t'$ and $\phi'$ as the coordinate values of $q$.
On the other hand, since the boundary null geodesics on the plane defined by $\theta_{1} = \theta_2 = \dots = \theta_{D-3} = \pi/2$ satisfy
\begin{align}
    \frac{dt}{d\phi}= \pm \ell,
\end{align}
the set of points on this plane  that are causally connected to $p$ in the sense of boundary causality is given by 
\begin{align}
t' \geq \ell |\phi'| \qquad (\phi' \in (- \pi, \pi]).
\label{t>|p|}
\end{align}
Therefore, if $q$ exists outside the region \eqref{t>|p|}, the null geodesic $\gamma(p,q)$ reaches a point earlier than the boundary causal curve. 
We say that such a null geodesic exhibits {\it time advance}.

\section{Time advance conditions for general static, spherically symmetric asymptotically AdS spacetime}
\label{sec:ta condition}

In this section, we derive the conditions for time advance in a general static, spherically symmetric asymptotically AdS spacetime, by solving the geodesic equations. 
The spherical symmetry allows us to assume that the geodesic is on the plane defined by $\theta_{1} = \theta_2 = \dots = \theta_{D-3} = \pi/2$ without loss of generality. 
Suppose a geodesic $\gamma(p,q)$ from $p\in\partial \mathcal{M}$ to $q\in\partial \mathcal{M}$ is a null geodesic that we consider here, and we set $t = \phi = 0$ at the point $p$.
Its orbit can be obtained by the integration of the geodesic equations. 
Let $\Delta t$ and $\Delta \phi$ be the increase of the coordinate value along the bulk null geodesic. 
Clearly, we obtain $t' = \Delta t$ but $\phi' \neq \Delta \phi$ generally because of $2 \pi$ times integer difference.
In this paper, we focus on the case\footnote{
If $\Delta \phi$ is negative, flipping the sign of $\phi$ ($\phi\to-\phi$)  results in a positive $\Delta \phi$. Moreover, since we perform a perturbative analysis to investigate time advance in this paper, time advance can only occur when $|\Delta\phi| = \pi +{\cal O} (\epsilon)$, where $\epsilon$ is defined after Eq.~\eqref{genep}. 
Therefore, the cases with $0 \leq \Delta \phi \leq \pi$ and $\pi < \Delta \phi< 2\pi$ cover all possibilities of the perturbative analysis. The case with $\pi < \Delta \phi< 2\pi$ is discussed in Appendix \ref{app:largephi} and gives the same results as those with $0 \leq \Delta \phi \leq \pi$.
} with $0 \leq \Delta \phi \leq \pi$. 
Then, since $\phi' = \Delta \phi$ holds,  Eq.~\eqref{t>|p|} gives the conditions for time advance
\begin{align}
    &\begin{cases}
        \Delta\phi - \pi \leq 0, \\
        \Delta t - \ell\Delta\phi < 0.
    \end{cases} \label{time advance conditions}
\end{align}
The first condition is an artificial restriction on $\phi$ introduced for analytical convenience, while the second condition is the actual statement of time advance. Therefore, these two conditions together constitute a sufficient condition for time advance. In this paper, however, we refer to them collectively as the {\it time advance conditions.}

In subsection \ref{subsec:AE}, focusing on null geodesics in the asymptotic region, we rewrite the time advance conditions \eqref{time advance conditions} in terms of the metric functions. 
Then, we provide a more detailed analysis of some special cases in subsections \ref{subsec:2term} and \ref{subsec:3term}.

\subsection{Asymptotic expansion}
\label{subsec:AE}
Let us consider the case where the effective metric is a general static, spherically symmetric asymptotically AdS spacetime,
\begin{align}
    \widehat{g}_{\mu\nu} dx^{\mu} dx^{\nu} = -f(r)dt^2+\frac{h(r)}{f(r)}dr^2 + g(r) r^2 d\Omega^2_{D-2}. \label{general}
\end{align}
Here, $f(r)$, $h(r)$ and $g(r)$ are functions of the radial coordinate $r$. 
Suppose that the metric functions are expressed as, 
\begin{align}
    &f(r)  =\frac{r^2}{\ell^2} + 1+\sum^\infty_{n=1}\frac{f_n}{r^n}, \label{fr}\\
    &h(r) = 1 + \sum^\infty_{n=1}\frac{h_n}{r^n}, \label{hr}\\
    &g(r) =  1 + \sum_{n=3}^{\infty} \frac{g_{n}}{r^{n}}, \label{gr}
\end{align}
where $f_n$, $h_n$, and $g_{n}$ are constants\footnote{
Note that $g_{1} = g_{2} = 0$ is assumed due to the asymptotic condition for asymptotically AdS spacetime with respect to the areal radius. 
}.

Let us focus on a future directed null geodesic $\gamma(p,q)$ from $p$ to $q$ associated with this effective metric.
Due to the spherical symmetry, the null geodesic can be restricted to that on the equatorial plane $\theta_1=\theta_2=...=\theta_{D-3}=\pi/2$.
Thus, the tangent vector of the null geodesic, say $k^{\mu}$, possesses only the $t$, $r$, and $\phi$ components, 
\begin{align}
 k^{\mu}\partial_{\mu}  = \dot{t}(\tau) \partial_{t} + \dot{r}(\tau) \partial_{r} + \dot{\phi}(\tau) \partial_{\phi},
\end{align}
where the dot denotes the derivative with respect to the affine parameter $\tau$. 
Note that $\dot{t}$ is positive so that the geodesic is future directed.
We focus on geodesics with $\dot{\phi} > 0$ without loss of generality.

We can define two conserved quantities $E$ and $L$ along the null geodesics by
\begin{align}
 E & \coloneqq - \widehat{g}_{\mu\nu} k^{\mu} (\partial_{t})^{\nu} = f(r) \dot{t}\, (>0) , \label{def:E}\\
 L &\coloneqq   \widehat{g}_{\mu\nu} k^{\mu} (\partial_{\phi})^{\nu} = g(r) r^2 \dot{\phi}\, (>0) \label{def:L},
\end{align}
associated with the Killing vectors $\partial/\partial t$ and $\partial/\partial\phi$ respectively.
Then, the null condition $g_{\mu\nu}k^{\mu} k^{\nu} = 0$ can be expressed as
\begin{align}
    \dot{r} = \pm E \sqrt{ \frac{1}{h(r)} \left[
    1-\frac{b^2f(r)}{r^2 g(r)} 
    \right]} ,\label{rdot=}
\end{align}
where we introduce the impact parameter $b$ by
\begin{align}
    b \coloneqq \frac{L}{E}.
\end{align}
The null geodesic $\gamma(p,q)$ of interest  starts at a point $p$ on the AdS boundary $\partial\mathcal{M}$ and ends at $q\in\partial \mathcal{M}$. 
Since both $p$ and $q$ exist at infinity $r\to\infty$, the geodesic has the minimum value of $r$. 
We denote the minimum by $r_m$, and refer the point that minimizes $r$ as the turning point $\lambda$. 
 $\dot{r} = 0$ holds at $\lambda$, and thus,
by setting $r = r_{m}$ in Eq.~\eqref{rdot=}, we can express the impact parameter $b$ as 
\begin{align}
b = r_m \sqrt{\frac{g(r_{m})}{f(r_m)}}.
\end{align}
Between $p$ and $\lambda$, $r$ is a decreasing function of $\tau$, with the minus sign in Eq.~\eqref{rdot=} corresponding to this region. 
In contrast, the plus sign in Eq.~\eqref{rdot=} corresponds to the region between $\lambda$ and $q$.

From Eqs.~\eqref{def:E} and~\eqref{def:L}, as well as Eq.~\eqref{rdot=},
we obtain
\begin{align}
 \frac{d t}{d r} &= \frac{\dot{t}(\tau)}{\dot{r}(\tau)} = \pm \frac{\sqrt{h(r)}}{f(r) \sqrt{1 - \frac{b^2 f(r)}{r^2 g(r)}}}, \\
 \frac{d \phi}{dr} &= \frac{\dot{\phi}(\tau)}{\dot{r}(\tau)} = \pm \frac{b}{r^2 g(r) } \frac{\sqrt{h(r)}}{\sqrt{1 - \frac{b^2 f(r)}{r^2 g(r)}}}.
\end{align}
Then, $\Delta t$ and $\Delta \phi$ along the boundary-to-boundary null geodesic $\gamma(p,q)$ can be evaluated as 
\begin{align}
   &\Delta t=2\int^\infty_{r_m} dr\frac{\sqrt{h(r)}}{f(r)\sqrt{1-\frac{b^2f(r)}{r^2 g(r)}}},\label{genet} \\
    &\Delta\phi=2\int^\infty_{r_m} dr\frac{b\sqrt{h(r)}}{r^2 g(r) \sqrt{1-\frac{b^2 f(r) }{r^2 g(r)}}} \label{genep}.
\end{align}
We can evaluate these integrals analytically under the approximations $f_n/r_m^n, h_n/r_m^n, g_n/r_m^{n} \ll 1$, denoting the order of them as $\mathcal{O}(\epsilon)$. As shown in Appendix~\ref{app:deltaphit}, $\Delta\phi - \pi$ and  $\Delta t/\ell-\Delta\phi$ can be evaluated as\footnote{
As shown in Appendix~\ref{app:largephi}, the results in the case with $\pi<\Delta \phi<2\pi$ coincide with those in the case with $\Delta \phi - \pi \leq 0$.}
\begin{align}
    \Delta\phi-\pi &= 
\sum_{n=1}^{\infty} \frac{C_{n}}{r_{m}^{n}} + \mathcal{O}(\epsilon^2),\label{timeadv1} \\
     \Delta t/\ell-\Delta\phi &= - \sum_{n=1}^{\infty} \beta_{n} \frac{C_{n}}{r_m^{n}} + \mathcal{O}(\epsilon^2),\label{timeadv2}
\end{align}
where the coefficients $C_{n}$ and $\beta_{n}$ are given by
\begin{align}
C_{n} &\coloneqq \frac{1}{2} B \left( \frac{1}{2}, \frac{1 + n}{2} \right) \Big(
    -(n + 1) f_n + h_n + \left(n - 1\right) g_{n} + (n + 1) \frac{g_{n + 2}}{\ell^2} 
    \Big), \notag\\
\beta_{n} &\coloneqq 1 - {}_{2}F_{1} \left( \frac{n}{2}, \frac{1}{2}, \frac{n}{2} + 1; - \frac{\ell^2}{r_{m}^2}\right),
\end{align}
where $B$ and ${}_{2}F_{1}$ are the beta function and the Gauss's hypergeometric function respectively. Here, we also impose the assumption that $C_{n}/r_m^n$ is of order $\mathcal{O}(\epsilon)$, which implies that, 
if we consider a geodesic with large $r_m/\ell$, then $g_{n + 2}$ must be significantly smaller than $\epsilon$ 
so that $g_{n + 2}/(\ell^2 r_m^n)$ is also of order $\mathcal{O}(\epsilon)$.
Properties of the factor $\beta_{n}$ are studied in Appendix~\ref{app:gamman}. 
Some of the key properties are 
\begin{align}
& 0 < \beta_{n} <  1,\label{gamma1}
\end{align}
and
\begin{align}
 & 1 < \frac{\beta_{n_{2}}}{\beta_{n_{1}}} < 3 \qquad \text{for } n_{1} < n_{2}.\label{gamma2}
\end{align}

With the results obtained above, we now present the time advance conditions in terms of $C_n$. 
The time advance conditions for the geodesic with the turning point $r_{m}$, namely, $\Delta \phi - \pi \leq 0$ and $\Delta t/\ell - \Delta \phi < 0$, can be summarized as
\begin{align}
 \mathcal{F} &\coloneqq \sum_{n = 1}^{\infty} \frac{C_{n}}{r_{m}^n} \leq 0, \label{def:calF}\\
 \mathcal{G} &\coloneqq \sum_{n=1}^{\infty} \beta_{n} \frac{C_{n}}{r_{m}^n} >0.\label{def:calG}
\end{align}
An immediate consequence from these inequalities is that no time advance occurs when all $C_{n}$'s are same signature, because either of these conditions is not satisfied.
Additionally, other necessary conditions can be obtained from the inequality 
\begin{align}
 \beta \mathcal{F} - \mathcal{G} < 0,\label{simlar1}
\end{align}
for arbitrary positive constant $\beta$.
If one choose $\beta$ as $\beta_{\bar{n}} \leq \beta < \beta_{\bar{n}+1}$, the inequality can be expressed as 
\begin{align}
\beta \mathcal{F} - \mathcal{G} = \sum_{n = 1}^{\bar{n}} (\beta - \beta_{n}) \frac{C_{n}}{r_{m}^{n}} - \sum_{\bar{n} + 1}^{\infty} (\beta_{n} - \beta) \frac{C_{n}}{r_{m}^{n}} < 0.
\end{align}
This inequality cannot be satisfied if all $C_{n}$ with $ n \leq \bar{n}$ are positive and all $C_{n}$ with $ n > \bar{n}$ are negative.
This requires the existence of an integer $n$ such that $C_n$ is negative and an integer $n'>n$ such that $C_{n'}>0$.

\subsection{Detailed Analysis for 2-terms case}
\label{subsec:2term}
In this subsection, we focus on the case where only two of $C_{n}$, precisely $C_{n_{1}}$ and $C_{n_{2}}$ with $n_{1} < n_{2}$, are nonzero.  
From the discussion in Sec.~\ref{subsec:AE}, the existence of a time advance null geodesic requires $C_{n_{1}} < 0$ and $C_{n_{2}} > 0$. 
The functions $\mathcal{F}$ and $\mathcal{G}$ are given by
\begin{align}
 \mathcal{F}(r_{m}) &= \frac{C_{n_{1}}}{r_{m}^{n_{1}}} + \frac{C_{n_{2}}}{r_{m}^{n_{2}}},  \\
 \mathcal{G}(r_{m}) &= \beta_{n_{1}} \frac{C_{n_{1}}}{r_{m}^{n_{1}}} + \beta_{n_{2}} \frac{C_{n_{2}}}{r_{m}^{n_{2}}}.
\end{align}
A geodesic with $r_{m}$ is time advanced if $r_{m}$ satisfies $\mathcal{F}(r_{m}) \leq 0$ and $\mathcal{G}(r_{m}) > 0$. These conditions can be expressed as 
\begin{align}
\left(\frac{C_{n_{2}}}{|C_{n_{1}}|} \right)^{\frac{1}{n_{2} - n_{1}}} \leq r_{m} < \left( \frac{\beta_{n_{2}}}{\beta_{n_{1}}} \frac{C_{n_{2}}}{|C_{n_{1}}|} \right)^{\frac{1}{n_{2} - n_{1}}}.\label{rmregiontwoterm}
\end{align}
Although $\beta_{n_{2}}/ \beta_{n_{1}}$ depends on $r_{m}$ in general, it always satisfies $\beta_{n_{2}} / \beta_{n_{1}} > 1$. Therefore, $r_{m}$ which satisfies the inequality \eqref{rmregiontwoterm}, {\it i.e.}, the time advance conditions, always exists for any given $C_{n_{1}} < 0$ and $C_{n_{2}} > 0$.

Let us evaluate the valid range of the approximations. 
The radial coordinate at the turning point of the would-be time advance null geodesic is 
\begin{align}
r_{m} \sim \left(C_{n_{2}}/|C_{n_{1}}| \right)^{\frac{1}{n_{2} - n_{1}}}. 
\label{rmc1c2}
\end{align}
In our analysis, we assume that the conditions $|C_{n_{1}}|/r_{m}^{n_{1}} \ll 1$ and $|C_{n_{2}}|/r_{m}^{n_{2}} \ll 1$ are satisfied. 
From Eq.~\eqref{rmc1c2}, these conditions require  
\begin{align}
\left(
\frac{|C_{n_{1}}|^{\frac{1}{n_{1}}}}{C_{n_{2}}^{\frac{1}{n_{2}}}}
\right)^{\frac{n_{1} n_{2}}{n_{2} - n_{1}}} \ll 1.\label{2terms1}
\end{align}
Thus our analysis is valid when the length scale of $C_{n_{1}}$ is much shorter than that of $C_{n_{2}}$. 

\subsection{Detailed Analysis for 3-terms case}
\label{subsec:3term}
In this subsection, we focus on the case where all $C_{n}$ except for $n = n_{1}, n_{2}, n_{3}$ vanish. Without loss of generality, we set $n_{1} < n_{2} < n_{3}$.
In this setup, the functions $\mathcal{F}$ and $\mathcal{G}$ are given as 
\begin{align}
 \mathcal{F} &= \frac{C_{n_{1}}}{r_{m}^{n_{1}}} + \frac{C_{n_{2}}}{r_{m}^{n_{2}}} + \frac{C_{n_{3}}}{r_{m}^{n_{3}}}, \notag\\
 \mathcal{G} &= \beta_{n_{1}} \frac{C_{n_{1}}}{r_{m}^{n_{1}}} + \beta_{n_{2}} \frac{C_{n_{2}}}{r_{m}^{n_{2}}} + \beta_{n_{3}} \frac{C_{n_{3}}}{r_{m}^{n_{3}}}.
\end{align}
We also assume $C_{n_{1}} \geq 0$, which corresponds to the positivity of the mass in the applications in the next section. 
Then, from the discussion in Sec.~\ref{subsec:AE}, we find that the time advance geodesic exists only when $C_{n_{2}} < 0$ and $C_{n_{3}} > 0$.

To see the behavior of these functions, it is useful to define new variable 
\begin{align}
x \coloneqq r_{m}^{n_{3} - n_{2}} > 0,
\end{align}
and rewrite the functions $\mathcal{F}$, $\mathcal{G}$  as
\begin{align}
 \widehat{\mathcal{F}}(x) &\coloneqq r_{m}^{n_{3}} \mathcal{F} = C_{n_{1}} x^{N} + C_{n_{2}} x + C_{n_{3}}, 
  \label{hatF}
\\
 \widehat{\mathcal{G}}(x) &\coloneqq r_{m}^{n_{3}} \mathcal{G} = \beta_{n_{1}} C_{n_{1}} x^{N} + \beta_{n_{2}} C_{n_{2}} x +  \beta_{n_{3}} C_{n_{3}},
 \label{hatG}
\end{align}
where we introduce the number $N \coloneqq \frac{n_{3} - n_{1}}{n_{3} - n_{2}} > 1$.
Now the time advance conditions can be expressed as $\widehat{\mathcal{F}} \leq 0$ and $\widehat{\mathcal{G}} > 0$.

Let us investigate the properties of the function $\widehat{\mathcal{F}}$. 
The positivity of $C_{n_{1}}$ and $N>1$ imply that $\widehat{\mathcal{F}}$ is a convex downward function of $x$.
The minimum value of $\widehat{\mathcal{F}}$ is achieved at $x=x_{\text{min}}\coloneqq(|C_{n_2}|/(NC_{n_1}))^{1/{N-1}} $ and it can be evaluated as 
\begin{align}
  \widehat{\mathcal{F}} (x_{\text{min}})\coloneqq \widehat{\mathcal{F}}_{\text{min}} = - \frac{N - 1}{N}\left(\frac{1}{N} \frac{|C_{n_{2}}|}{C_{n_{1}}}\right)^{\frac{1}{N-1}} |C_{n_{2}}| + C_{n_{3}}. 
\end{align}
Therefore, $\widehat{\mathcal{F}}$ has zero point(s) only when the minimum $\widehat{\mathcal{F}}_{\text{min}}$ is non-positive, {\it i.e.},
\begin{align}
\frac{N^N}{(N-1)^{N-1}} \left( \frac{C_{n_{1}} C_{n_{3}}^{N - 1}}{|C_{n_{2}}|^{N}}\right)  \leq 1 .\label{necessaryconditionforF<0}
\end{align}
Then, one of the time advance condition $\widehat{\mathcal{F}} \leq 0$ can be expressed as
$ x_{1} \leq x \leq x_{2}$, letting the zero points of $\widehat{\mathcal{F}}$ be $x_{1}$ and $x_{2}$ with $x_{1} \leq x_{2}$.

To proceed with the analysis, we consider the case where the $C_{n_1}$ terms in Eqs.~\eqref{hatF} and \eqref{hatG} are negligible\footnote{ 
In this approximation, the positivity of $C_{n_1}$ is not essential to the result, although $x_2$ may not exist for $C_{n_1}\leq 0$.
}. 
This case reduces to the $2$-terms case discussed in the previous subsection. 
Then, the smaller solution of $\widehat{\mathcal{F}} (x) =0$
is $x_{1} \simeq {C_{n_{3}}}/{|C_{n_{2}}|}$, and at $x=x_1$ 
the first term and the other two terms in the right hand side of Eq.~\eqref{hatF} are 
estimated as $C_1 ({C_{n_{3}}}/{|C_{n_{2}}|})^N$ and $C_{n_{3}}$, respectively. 
Hence, the condition for the first term to be negligible is expressed by
\begin{align}
 \epsilon_{C} \coloneqq \frac{C_{n_{1}} C_{n_{3}}^{N - 1}}{|C_{n_{2}}|^{N}} \ll 1.
\label{ec}
\end{align}
In this case, we can express $x_{1}$ and $x_{2}$ analytically.
The contribution of $C_{n_{1}}$ term is negligible around $x_{1}$, while the detailed analysis shows that the $C_{n_{3}}$ term is negligible near $x_{2}$.
Solving $\widehat{\mathcal{F}} = 0$ in this approximation, we obtain,
\begin{align}
x_{1} &= \frac{C_{n_{3}}}{|C_{n_{2}}|} \left( 1 + \mathcal{O} \left( \epsilon_{C} \right) \right), \\
x_{2} &= \left( \frac{|C_{n_{2}}|}{C_{n_{1}}} \right)^{\frac{1}{N-1}} \left( 1 + \mathcal{O} \left( \epsilon_{C}^{\frac{1}{N-1}} \right) \right).
\end{align}

Similarly, $C_{n_{1}}$ term in $\widehat{\mathcal{G}}$ is negligible near the smallest zero point, say $x = x_{3}$ \footnote{
More precisely, ${\beta_{n_1}C_{n_{1}} (\beta_{n_3} C_{n_{3}})^{N - 1}}/({\beta_{n_2}|C_{n_{2}}|)^{N}} \ll 1$ is required for this approximation. 
Since $\beta_{n_1}\beta_{n_3}^{N-1}/\beta_{n_2}^N$ could be enormous for large $N$, this approximation might be different from Eq.~\eqref{ec}. 
However, because the analysis here is done after fixing $n_i$'s, 
we denote $\beta_{n_1}\beta_{n_3}^{N-1}/\beta_{n_2}^N$ as a value of order unity.
}.
Under this approximation, $x_{3}$ can be obtained as 
\begin{align}
  x_{3} &= \frac{\beta_{n_{3}}}{\beta_{n_{2}}}\frac{C_{n_{3}}}{|C_{n_{2}}|} ( 1 + \mathcal{O}(\epsilon_{C})).
\end{align}
The condition $\widehat{\mathcal{G}} > 0$ can be satisfied for $0 < x < x_{3}$.
Since the factor $\beta_{n_{3}}/ \beta_{n_{2}}$ satisfies $ 1 < \beta_{n_{3}}/ \beta_{n_{2}} < 3$, we obtain $x_{1} < x_{3} \ll x_{2}$. 
Therefore the time advance condition can be satisfied for a geodesic with 
\begin{align}
 x_{1} < x < x_{3}.
\end{align}
Translating into the condition for $r_{m}$, we obtain
\begin{align}
 \left( \frac{C_{n_{3}}}{|C_{n_{2}}|} \right)^{\frac{1}{n_{3} - n_{2}}} \leq r_{m} < \left( \frac{\beta_{n_{3}}}{\beta_{n_{2}}} \frac{C_{n_{3}}}{|C_{n_{2}}|} \right)^{\frac{1}{n_{3} - n_{2}}}.\label{3terms1}
\end{align}
This result is consistent with the 2-terms analysis in the previous subsection.

Before moving to the concrete applications of our formula, let us summarize the approximation used here. 
Since the radial coordinate of the turning point of the time advance null geodesic is the order of $(C_{n_{3}}/|C_{n_{2}}|)^{1/(n_{3} - n_{2})}$, we can express the condition for the approximation in terms of the $C_{n_{i}}$.
First, the conditions for $\epsilon \ll 1$, which are used to derive Eqs.~\eqref{timeadv1} and~\eqref{timeadv2}, can be expressed as 
\begin{align}
\frac{|C_{n_{i}}|}{r_{m}^{n_{i}}} \sim 
 \frac{|C_{n_{i}}| |C_{n_{2}}|^{\frac{n_{i}}{n_{3} - n_{2}}}}{C_{n_{3}}^{\frac{n_{i}}{n_{3} - n_{2}}}} \eqqcolon  \epsilon_{n_{i}}  \ll 1.\label{3term2}
\end{align}
More explicitly, $\epsilon_{n_{i}}$ can be expressed as
\begin{align}
\epsilon_{n_{1}} &= C_{n_{1}} \left( \frac{ |C_{n_{2}}|}{C_{n_{3}}} \right)^{\frac{n_{1}}{n_{3}- n_{2}}} \ll 1, \\
\epsilon_{n_{2}} &= \epsilon_{n_{3}} = \left( \frac{|C_{n_{2}}|^{1/n_{2}}}{C_{n_{3}}^{1/n_{3}}} \right)^{\frac{n_{2} n_{3}}{n_{3} - n_{2}}} \ll 1.
\label{en2en3}
\end{align}
Second, the condition $\epsilon_{C} \ll 1$, which is used to derive the analytical expressions for the zero points for the functions $\mathcal{F}$ and $\mathcal{G}$, is
\begin{align}
 \epsilon_{C} = C_{n_{1}}  \frac{C_{n_{3}}^{N - 1}}{|C_{n_{2}}|^{N}} 
= \frac{C_{n_{1}} C_{n_{3}}^{\frac{n_{2}-n_{1}}{n_{3}-n_{2}}}}{|C_{n_{2}}|^{\frac{n_{3}-n_{1}}{n_{3}-n_{2}}}}
\ll 1.\label{3term3}
\end{align}
One can check that the relation $\epsilon_{n_{1}} = \epsilon_{n_{2}} \epsilon_{C}$ holds. Hence $\epsilon_{n_{1}} \ll 1$ is automatically satisfied if $\epsilon_{n_{2}} \ll 1$ and $\epsilon_{C} \ll 1$ are satisfied.

\section{Applications}
\label{sec:application}

Using the formulas which are obtained in the previous section, 
we investigate the possibilities of time advance in the minimally coupled Einstein--Maxwell theory with the negative cosmological constant (Sec.~\ref{subsec:EM}) and 
in the Einstein--Euler--Heisenberg theory (Sec.~\ref{subsec:EH}). 

\subsection{Einstein--Maxwell Theory}
\label{subsec:EM}

In this subsection, we explore the static, spherically symmetric solution in the minimally coupled Einstein--Maxwell theory with the negative cosmological constant, namely the Reissner--Nordstr\"{o}m anti-de Sitter (RNAdS) spacetime.
The action of the Einstein--Maxwell theory in $D$-dimensional spacetime is given by
\begin{align}
 S = \int d^{D} x \frac{\sqrt{-g}}{\Omega_{D-2}} \left[ \frac{1}{2 (D-2) G} \left(R - 2 \Lambda \right) - \frac{1}{4 k} F_{\mu\nu} F^{\mu\nu} \right],
\end{align}
where $G$ and $k$ are the gravitational coupling constant and the Coulomb constant respectively, and $\Omega_{D-2}$ is the volume of the unit $D-2$ dimensional sphere given by $ \Omega_{D-2}\coloneqq 2\pi^{\frac{D-1}{2}}/\Gamma\left(\frac{D-1}{2}\right)$.
The cosmological constant $\Lambda$ is related with the AdS radius $\ell$ through
\begin{align}
 \Lambda = - \frac{(D-1)(D-2)}{2} \frac{1}{\ell^2}.
\end{align}

As is well known, the characteristic surfaces of the Einstein--Maxwell equation are generated by the null geodesics with respect to the spacetime metric $g_{\mu\nu}$. Thus, the effective metric $\widehat{g}_{\mu\nu}$, shown in Eq.~\eqref{general}, is simply the spacetime metric $g_{\mu\nu}$.

The static, spherically symmetric solution of this system is the Reissner--Nordstr\"{o}m anti-de Sitter (RNAdS) spacetime, which corresponds to the metric \eqref{general} with
\begin{align}
    &f(r)= 1 + \frac{r^2}{\ell^2}-\frac{2GM}{r^{D-3}}+\frac{1}{D-3}\frac{G k Q^2}{r^{2(D-3)}}, \label{qwe1} \\
    &h(r)=1,\label{qwe1h}\\
    &g(r)=1, \label{qwe1g}
\end{align}
as well as the Coulomb potential,
\begin{align}
 A_{\mu} dx^{\mu} = - \Phi(r) dt,
\end{align}
with
\begin{align}
 \Phi(r) = \frac{k}{D-3} \frac{Q}{r^{D-3}}.
\end{align}
Since the characteristic surface is generated by the null geodesics with respect to the spacetime metric in the Einstein--Maxwell theory, we can apply the general analysis in the previous section by setting 
\begin{align}
    f_{D-3} = - 2 G M, \qquad f_{2(D-3)} = \frac{k G Q^2}{D-3},\label{highRNAdS}
\end{align}
and other $f_n$, $h_n$ and $g_{n}$ vanish.
Then, the parameters $C_{n}$ can read as
\begin{align}
 C_{D-3} &= (D - 2) B\left(\frac{1}{2}, \frac{D - 2}{2}\right) G M, \\
 C_{2(D-3)} &= - \frac{2 D - 5}{2 (D - 3)} B\left(\frac{1}{2}, \frac{2 D - 5}{2}\right)  G k Q^2,
\end{align}
and other $C_{n}$ vanish. Since the coefficient $C_{2(D - 3)}$ is negative, no geodesic satisfies the time advance conditions.

We would like to emphasize that the sub-extremal condition is not assumed, as well as the positivity of the ADM mass $M$. 
Thus, our result shows that no time advance occurs in the spacetime region $(G |M|)^{1/(D-3)} \ll r$ and $(G k Q^2)^{1/2(D-3)} \ll r$ even for the spacetime with naked singularities\footnote{
Our analysis applies to the large $r$ region of the super-extremal solution. 
The presence of a naked singularity is not required, and such configurations naturally arise, for example, when an electron is located at the center.
}. This situation is beyond the general discussion by Gao and Wald \cite{Gao:2000ga}, where the global hyperbolicity\footnote{
By globally hyperbolic, we mean the condition that combines strong causality with the compantness of $J^{+}(p) \cap J^{-}(q)$ for any point $p,q \in \overline{M}$, following the standard definition found in textbooks such as Hawking and Ellis~\cite{Hawking:1973uf}.
} is assumed.

\subsection{Einstein--Euler--Heisenberg Theory}\label{subsec:EH}

In this subsection, we will explore the time advance condition in the Einstein--Euler--Heisenberg type of the effective field theory, where the Lagrangian is given by
\begin{align}
	S=\int d^D x \frac{\sqrt{-g}}{\Omega_{D-2}}
	&\Bigg[
	\frac{1}{2 (D-2) G}(R-2\Lambda)-\frac{1}{4 k} F^{\mu\nu}F_{\mu\nu} + \alpha_1 F_{\mu\nu}F^{\mu\nu}F_{\rho\sigma}F^{\rho\sigma}+\alpha_2 F^{\mu\nu}F^{\rho\sigma}F_{\nu\rho}F_{\sigma\mu}
	\Bigg].\label{Action EH}
\end{align}
From the general perspective of the effective field theory approach, our analysis corresponds to neglecting the interactions that include spacetime curvatures among the fourth-order derivative terms in the derivative expansions, such as $R_{\mu\nu\rho\sigma} F^{\mu\nu} F^{\rho\sigma}$, $R_{\mu\nu\rho\sigma} R^{\mu\nu\rho\sigma}$.

Since the theory with the Lagrangian \eqref{Action EH} is regarded as the leading order terms in the derivative expansions, 
we treat the corrections from the $\alpha_{1}$ and $\alpha_{2}$ terms perturbatively. This treatment can be justified when the $\alpha_{1}$ and $\alpha_{2}$ terms are sufficiently small, compared to the Einstein--Maxwell terms in the action.
As derived in Eq.~\eqref{validity123} of the appendix~\ref{App:bgsolution}, this requires either
\begin{align}
\varepsilon_{\alpha}& \sim  \frac{k^2 |\alpha_{i}|}{G r^2} \epsilon \ll 1 \qquad \mbox{or} \qquad \frac{\ell^2}{r^2} \frac{k^2 |\alpha_{i}|}{G r^2} \epsilon^2 \ll 1 .
\end{align}
In addition, the static, spherically symmetric solution includes the correction from the Reissner--Nordstrom solution only in the order $\mathcal{O}(\epsilon \varepsilon_{\alpha})$. 
Therefore, we can use the Reissner--Nordstrom solution as the background spacetime $\bar{g}_{\mu\nu}$ in the leading order analysis in $\mathcal{O}(\epsilon)$ or $\mathcal{O}(\varepsilon_{\alpha})$.

Since no kinetic structures for gravitons are modified by Euler--Heisenberg correction terms, the characteristic surface for the graviton propagation is governed by simply the spacetime metric $\bar{g}_{\mu\nu}$.
Hence the analysis for the gravitational wave is same as the Einstein--Maxwell case. On the other hand, the characteristic surfaces of the electromagnetic wave include the corrections due to the terms proportional to $\alpha_i$. 
In order to obtain the effective metric of the electromagnetic wave, we perform the mode decomposition of it.
The analysis in Appendix \ref{App:emp} shows that the effective metrics for the scalar and the vector modes are given as
\begin{align}
 ( \widehat{g}~^{-1})^{\mu\nu} = \bar{g}^{\mu\nu} - 8 k \alpha_{A} F_{\rho}{}^{\mu} F^{\rho\nu},
\end{align}
where $A =\{S, V\}$ with $\alpha_S = 4\alpha_1 + 2\alpha_2$ for the scalar mode and $\alpha_V = \alpha_2$ for the vector mode
\footnote{
Our expression for the effective metrics reproduces the result in Ref.~\cite{DeLorenci:2000yh} in the case of $D=4$ flat spacetime.
}
.
Then, substituting the components of the spacetime metric $\bar{g}_{\mu\nu}$ and the Coulomb potential $\bar{\Phi}$, we obtain
\begin{align}
    f(r) &= 
    1+\frac{r^2}{\ell^2}-\frac{2GM}{r^{D-3}} + \frac{1}{D - 3} \frac{G k Q^2}{r^{2(D-3)}}
    \nonumber\\
    &\qquad 
    -\frac{8 k^3 Q^2 \alpha_A}{\ell^2r^{2(D-3)}}
-\frac{8 k^3 Q^2 \alpha_A}{r^{2(D-2)}} + \mathcal{O}(\epsilon \varepsilon_{\alpha}, \epsilon^2, \varepsilon_{\alpha}^2), \label{EHf}\\ 
    h(r) &= 1 - \frac{16 k^3 Q^2 \alpha_A}{r^{2(D-2)}} + \mathcal{O}(\epsilon \varepsilon_{\alpha}, \epsilon^2, \varepsilon_{\alpha}^2) ,\label{EHh}\\ 
    g(r) &= 1 + \mathcal{O}(\epsilon \varepsilon_{\alpha}, \epsilon^2, \varepsilon_{\alpha}^2) ,
\end{align}
where $\mathcal{O}(\epsilon)$ denotes the order of $GM/r^{D-3}$, $k G Q^2/r^{2(D-3)}$, while $\varepsilon_{\alpha}$ denotes the order of $k^3 Q^2 \alpha_{i}/r^{2(D-2)}$.
In terms of the notations in Eqs.~\eqref{fr}--\eqref{gr}, the non-zero coefficients can be read as
\begin{align}
 f_{D-3} &= - 2 G M, \\
f_{2(D-3)} &= \frac{G k Q^2}{D - 3} - 8 \ell^{-2} k^3 Q^2\alpha_{A}, \\
f_{2(D-2)} &= - 8 k^3 Q^2 \alpha_{A},\\
h_{2(D-2)} &= - 16 k^3 Q^2  \alpha_{A}.
\end{align}
From these coefficients, we can evaluate non-vanishing $C_{n}$ as
\begin{align}
 C_{D-3} &= (D-2) B\left(\frac{1}{2}, \frac{D-2}{2}\right) G M, \\
 C_{2(D-3)} &= - \frac{2 D - 5}{2(D-3)} B\left( \frac{1}{2}, \frac{2D - 5}{2} \right) G k Q^2 \left( 1
 - 8 (D-3) \frac{ k^2 \alpha_{A}}{\ell^2 G} 
\right), \\
 C_{2(D-2)} &= 4 (2 D - 5) B\left(\frac{1}{2}, \frac{2 D - 3}{2}\right)
  k^3 Q^2 \alpha_{A}.
\end{align}
Thus, the system reduces to the 3-terms case with $n_{1} = D - 3, n_{2} = 2(D-3)$ and $n_{3} = 2 (D -2)$.

As a consequence of the discussion in the previous section, provided that $M \geq 0$, the time advance possibly occurs only when both $C_{2(D-3)} < 0$ and $C_{2 (D -2)} > 0$ hold. 
These conditions are expressed as
\begin{align}
 0 < \alpha_{A} < \frac{\ell^2 G}{8 (D - 3) k^2}.\label{timeadvancealpha}
\end{align}
More exactly, the conditions for the scalar and the vector modes are
\begin{align}
    &0 < 2\alpha_1+\alpha_{2} < \frac{\ell^2 G}{16 (D - 3) k^2},\label{ta condition:alpha1+2}\\
    &0 < \alpha_{2} < \frac{\ell^2 G}{8 (D - 3) k^2}\label{ta condition:alpha2}.
\end{align}
Then, as long as the approximations \eqref{3term3} and \eqref{en2en3}
 are valid, there is a time advance null geodesic with the turning point
\begin{align}
 r_{m} \sim \left( \frac{C_{n_{3}}}{|C_{n_{2}}|} \right)^{\frac{1}{n_{3} - n_{2}}} = \sqrt{\frac{2D-5}{2D-4}} \ell K^\frac{1}{2}\sim \ell K^\frac{1}{2},
 \label{k1/2}
\end{align} 
with
\begin{align}
 K \coloneqq \frac{8 (D - 3) \frac{k^2 \alpha_{A}}{\ell^2 G}}{1 - 8 (D - 3) \frac{k^2 \alpha_{A}}{\ell^2 G}}.
\end{align}
The small parameters $\epsilon_{n_{2}}$, $\epsilon_{n_{3}}$ and $\epsilon_{C}$ of the perturbative expansion appearing in Eqs. \eqref{3term3} and \eqref{en2en3}
are expressed as\footnote{
We aim to find a solution where a time advance exists for an arbitrary fixed $\alpha$.  
Hence, $M$ and $Q$ can be determined once $\alpha$ is fixed.  
Depending on $\alpha$, we can choose sufficiently small values of $M$ and $Q$ such that $\epsilon_{n_2}$ and $\epsilon_{C}$ are small.}
\begin{align}
&\epsilon_{n_{2}} \sim \epsilon_{n_{3}} \sim \frac{G k Q^2}{\ell^{2 (D - 3)}} \frac{k^2 \alpha_{A}}{\ell^2 G}K^{-(D - 2)}, \\
& \epsilon_{C}  \sim \frac{G M}{\ell^{D-3}} \left( \frac{G k Q^2}{\ell^{2(D-3)}}\right)^{-1} \left( \frac{k^2 \alpha_{A}}{G \ell^2} \right)^{-1} K^{\frac{D - 1}{2}}.
\label{eC}
\end{align}
For the validity of the derivative expansion, $\varepsilon_{\alpha}$ should be small, which is written with $\alpha_{i}$ as
\begin{align}
 \varepsilon_{\alpha} \sim \frac{k^2 |\alpha_{i}|}{G r_{m}^2} \epsilon_{n_i} \sim  \left( 1 - 8 (D - 3) \frac{k^2 \alpha_{A}}{G \ell^2}\right) \epsilon_{n_i}.
\end{align}
This implies that the smallness of $\varepsilon_{\alpha}$ is ensured by that of $\epsilon_{n_i}$.

For a given theory, that is, for given parameters $\ell, \alpha_{1},\alpha_2$ as well as $G, k$ and $D$, one can always consider a solution which satisfies $\epsilon_{n_{2}} \ll 1$ by considering sufficiently small $Q$.
Then, $\epsilon_{C} \ll 1$ can also be satisfied if one consider sufficiently small $M$. For such parameters, we can apply the analytic result given in the previous section and conclude the existence of time advance null geodesics. Thus, we found that for any parameter $\alpha_{i}$ which satisfies the conditions \eqref{ta condition:alpha1+2} and~\eqref{ta condition:alpha2}, there always exists a choice of a solution of the equations of motion derived from the Lagrangian \eqref{Action EH} which possesses time advance null geodesics.
As a result, if one requires that the Einstein--Euler--Heisenberg type of the effective field theory does not possess any time advance null geodesic for any solution, the allowed parameter region for $\alpha_{1}$ and $\alpha_2$ should be excluded from the conditions \eqref{ta condition:alpha1+2} and~\eqref{ta condition:alpha2}.

Let us examine the property of the solution that possesses time advance geodesics in more detail.
First, we can find that $G M^2$ must be much smaller than $k Q^2$ because 
\begin{align}
 (D - 3) \frac{G M^2}{k Q^2} = \epsilon_{C}^2 \epsilon_{n_{2}} \left(1 - 8 (D - 3) \frac{k^2 \alpha_{A}}{G \ell^2}\right) \ll 1.
\end{align}
This means that the solution with time advance geodesics is superextremal because
\begin{align}
 - g_{tt} &= \left( 1 - \frac{G M}{r^{D-3}}\right)^2 + \frac{r^2}{\ell^2} + \frac{G}{r^{2 (D-3)}} \left( \frac{1}{D - 3} k Q^2 - G M^2 \right) \notag\\
& \sim \left( 1 - \frac{G M}{r^{D-3}}\right)^2 + \frac{r^2}{\ell^2} + \frac{G k Q^2}{(D - 3) r^{2 (D-3)}} > 0,
\end{align}
and hence there is no apparent horizon.
Next, we can find that time advance geodesics are passing through the region with negative quasi--local energy.
To see this property, let us investigate a generalized Misner--Sharp quasi--local energy $m_{MS}$ defined in Refs.~\cite{PhysRev.136.B571,10.1143/PTP.63.1217,Hayward:1994bu, Nakao:1995ks, Maeda:2006pm,Maeda:2007uu, Maeda:2012fr}. 
In the Reissner--Nordstr\"{o}m--anti de Sitter solution \eqref{qwe1}--\eqref{qwe1g}, this local energy for an $r$--constant surface is written as
\begin{align}
m_{MS} (r) \coloneqq - \left[f -\left(1 + \frac{r^2}{\ell^2} \right) \right] \frac{r^{D-3}}{2G} = M - \frac{kQ^2}{2(D-3)r^{D-3}}.
\end{align}
Using the expression for the radial coordinate $r_m$ of the time advance geodesic estimated in Eq.~\eqref{k1/2},
the generalized Misner--Sharp quasi--local energy for the surface with this radial coordinate $r_m$ can be evaluated as 
\begin{align}
m_{MS} (r_m) & \sim
- \frac{k Q^2}{2 (D-3) \ell^{D-3} K^{\frac{D-3}{2}}}
\left( 1 - \frac{1}{4} \left(1 - 8 (D-3) \frac{k^2 \alpha_{A}}{G \ell^2} \right) \epsilon_{C} \right).
\label{mMS}
\end{align}
The smallness of $\epsilon_{C}$ implies that the last term on the right hand side of Eq.~\eqref{mMS} is negligible compared to the first term, meaning that the generalized Misner--Sharp quasi--local energy $m_{MS}(r_m)$ is negative. This indicates that the necessity of the negative energy, or, in other words, the violation of the weak energy condition, to realize this solution as the exterior solution of some matter source with the regular center, instead of the naked singularity.

We now examine the relationship between the occurrence of time advance and the energy conditions.
Since no specific assumptions are imposed on the central region of the spacetime—where, for instance, naked singularities may exist—it is not necessarily the case that violations of energy conditions, as implied by the Gao–Wald theorem, must occur. Nonetheless, given that such violations are often associated with the emergence of time advances, it is important to analyze this connection in detail.
In what follows, we investigate this relationship from the perspectives of both the physical and effective metrics.

On a static spherically symmetric spacetime with metric 
\begin{eqnarray}
g_{\mu\nu} dx^{\mu} dx^{\nu} = -f(r)\,dt^2 + \frac{h(r)}{f(r)}\,dr^2 + r^2 d\Omega^2, 
\end{eqnarray}
where, for simplicity, we set $g(r)=1$,
the Einstein tensor can then be expressed as
\begin{eqnarray}
G_{\mu\nu} dx^{\mu} dx^{\nu} = \rho(r) f(r)\,dt^2 + \frac{p_r(r)h(r)}{ f(r)}\,dr^2 + p_\theta(r)\,r^2 d\Omega^2,
\end{eqnarray}
with
\begin{eqnarray}
&&\rho =
\frac{(D - 3)(D - 2)}{2r^2} \left( 1 - \frac{f(r)}{h(r)} \right)
+ \frac{D - 2}{2r} \left( -\frac{f'(r)}{h(r)} + \frac{f(r) h'(r)}{h(r)^2} \right) ,\\
&&p_r =
\frac{(D - 3)(D - 2)}{2r^2} \left( -1 + \frac{f(r)}{h(r)} \right)
+ \frac{D - 2}{2r} \cdot \frac{f'(r)}{h(r)}, \\
&&p_\theta =
\frac{(D - 3)(D - 4)}{2r^2} \left( -1 + \frac{f(r)}{h(r)} \right)
+ \frac{D - 3}{2r} \left( \frac{2f'(r)}{h(r)} - \frac{f(r) h'(r)}{h(r)^2} \right)
+ \frac{f''(r)}{2h(r)} - \frac{f'(r) h'(r)}{4h(r)^2}. \nn
\end{eqnarray}
From these expressions, we obtain
\begin{eqnarray}
&&\rho + p_r = \frac{D - 2}{2r} \cdot \frac{f(r) h'(r)}{h(r)^2}, \label{eqrpr}\\
&&\rho + p_\theta =\frac{D - 3}{r^2} \left(1 - \frac{f(r)}{h(r)}\right)  + \frac{D - 4}{2r} \cdot \frac{f'(r)}{h(r)} + \left[
\frac{f''(r)}{2h(r)}
+ \frac{f(r) h'(r)}{2r h(r)^2}
- \frac{f'(r) h'(r)}{4 h(r)^2}
\right].\nn
\label{eqrpt}
\end{eqnarray}
The null energy condition, together with the Einstein equation, implies the null convergence condition, which requires that these quantities be non-negative.
In light of the Gao–Wald theorem, it is more appropriate to examine whether the null convergence condition, which is a condition on the geometry rather than on the energy-momentum content, holds for both the physical metric and the effective metric, rather than focusing solely on the null energy condition. In the following, we investigate both cases in detail.

We now consider the null convergence conditions for the physical metric of the Reissner–Nordström–anti-de Sitter spacetime with the Euler–Heisenberg correction, given in Eqs.~\eqref{fbarEH}-\eqref{gbarEH}. 
Since $h(r)$ is unity, Eq.\eqref{eqrpr} implies that $\rho + p_r$ vanishes. 
Furthermore, Eq.\eqref{eqrpt} reduces to
\begin{eqnarray}
\rho + p_\theta = \frac{GkQ^2}{r^{2(D-2)}} \left[ 
(D-2) - \frac{8(3D^2-13D+16)}{3D-7} (2 \alpha_1+ \alpha_2) \frac{k^3 Q^2}{r^{2(D-2)}}
\right]
\end{eqnarray}
Since the second term in the square brackets is of order ${\cal O}(\varepsilon_\alpha)$, 
the first term is dominant, and thus, $\rho + p_\theta$ is non-negative.
We therefore conclude that the physical metric satisfies the null convergence conditions.

Next we examine the null convergence conditions for the effective metric in the Reissner–Nordström–anti-de Sitter spacetime with the Euler–Heisenberg correction, where
$f(r)$ and $h(r)$ are given in Eqs.~\eqref{EHf} and \eqref{EHh}, respectively.
 In the present context, these quantities take the form:
\begin{eqnarray}
&&\rho + p_r = 16(D - 2) \left( 1 + \frac{r^2}{\ell^2} \right) \cdot \frac{k^3 Q^2 \alpha_A}{r^{2(D - 1)}} 
+ 
\mathcal{O}(\epsilon \varepsilon_{\alpha}, \epsilon^2, \varepsilon_{\alpha}^2),
\label{NECr}
\\
&&\rho + p_\theta = (D - 2) \frac{GkQ^2} {r^{2(D - 2)}}
\left( 
1 - \frac{8(D - 3)k^2 \alpha_A}{\ell^2 G}
\right)
\left(
1 - \frac{D^2 - 2D - 1}{(D - 2)(D - 3)} \cdot \frac{K \ell^2}{r^2}
\right)
\nn&& \hspace{100mm}
+ 
\mathcal{O}(\epsilon \varepsilon_{\alpha}, \epsilon^2, \varepsilon_{\alpha}^2).
\label{NECtheta}
\end{eqnarray}
It is of particular interest to investigate whether any null geodesics exhibiting time advance traverse regions in which the null convergence condition is violated.
We have shown that the time advance null geodesic exists if $\alpha_A$ satisfies the inequality~\eqref{timeadvancealpha}.
Under this condition, such geodesics pass through the minimum value $r_m$ of $r$
\begin{eqnarray}
r_{\min}^{(m)}(D) \leq r_m < r_{\max}^{(m)} \left. \left( D, w \right)\right|_{w=-\ell^2/r_m^2} ,
\end{eqnarray}
where the bounds are given by
\begin{eqnarray}
&&r_{\min}^{(m)}(D):= \sqrt{\frac{2D - 5}{2(D - 2)} \ell^2 K}, \\
&&
r_{\max}^{(m)}(D,w) :=   \sqrt{\frac{\beta_{2 (D - 2)}(w)}{\beta_{2 (D - 3)}(w)} \frac{2D - 5}{2(D - 2)} \ell^2 K}.
\end{eqnarray}
Equation~\eqref{NECr} with the inequality~\eqref{timeadvancealpha} ensures the positivity of $\rho + p_r$. 
The sign of $\rho + p_\theta$ is not immediately obvious, and  requires closer inspection of the final parenthetical term in Eq.\eqref{NECtheta}.
Since this term is a monotonically increasing function of $r (>0)$ and all other multiplicative factors are positive, it suffices to evaluate the expression at the upper bound of 
$r_m$, namely $r=r_{\max}^{(m)}(D,w)$. 
According to the numerical results presented in  Appendix~\ref{secb/b}, the following inequality holds:
\begin{eqnarray}
r_{\max}^{(m)}\left( D, w \right) < r_{\max}^{(m)}(D, 0)= \sqrt { \frac{(D - 2)^2}{(D - 1)(D - 3)} \cdot \frac{2D - 5}{2(D - 2)} \ell^2 K}.
\label{b/b}
\end{eqnarray}
It then follows that
\begin{eqnarray}
&&\left. \rho + p_\theta \right|_{r_m = r_{\max}^{(m)}(D,w), w= -\ell^2/r_m^2} 
< 
\left. \rho + p_\theta \right|_{r_m = r_{\max}^{(m)}(D,0)} \nn 
&&\hspace{5mm}
=
\frac{(D - 2)}{r^{2(D - 2)}} GkQ^2 
\left( 
1 - \frac{8(D - 3)k^2 \alpha_A}{\ell^2 G}
\right)
\left(
 1 - \frac{2(D^2 - 2D - 1)}{2D - 5} \cdot 
\frac{(D - 1)}{(D - 2)^2}
\right) <0.\nn
\end{eqnarray}
This inequality confirms that $\rho + p_\theta$ at $r=r_{\max}^{(m)}(D,w)$
becomes negative, indicating that the time-advanced null geodesic passes through the region where the null convergence condition is violated.

We therefore conclude that, in the Reissner–Nordström–anti-de Sitter spacetime with the Euler–Heisenberg correction,
all null geodesics exhibiting time advance pass through regions where the null convergence condition for the effective metric is violated,
whereas the null convergence condition for the physical metric remains satisfied.

\section{Summary and Discussion}
\label{sec:summary}
In this paper, we investigate the conditions for the existence of time advance null geodesics in a general static, spherically symmetric, asymptotically AdS spacetime, provided that the bulk causality is characterized by the effective metric.
Under the approximations $f_n/r_m^n,h_n/r_m^n,g_n/r_m^n\ll1$, we obtain the sufficient conditions for a null geodesic with  the radial coordinate $r_{m}$ at the turning point to be time advance, which are given by Eqs.~\eqref{def:calF} and~\eqref{def:calG}.
Then, we focus on two specific cases; one is the 2-terms case where only two of $C_n$ are present, and the other is the 3-terms case where only three of $C_n$ exist.
For the former case, we demonstrate that, if the coefficients satisfy $C_{n_{1}} < 0$ and $C_{n_2}>0$, with the condition for the validity of approximation \eqref{2terms1}, there always exists a time advance null geodesic whose turning point $r_{m}$ is located in the region \eqref{rmregiontwoterm}. 
For the latter case, assuming that $C_{n_{1}} \geq 0$ holds, which corresponds to the positivity of the mass $M$ in Sec.~\ref{sec:application},
we find that if the other two coefficients satisfy $C_{n_{2}} < 0$ and $C_{n_{3}} > 0$ along with the additional assumption  \eqref{ec} to simplify the analysis, as well as the condition for the validity of the approximations \eqref{en2en3}, there always exists a time advance null geodesic located in the region~\eqref{3terms1}.

Then, we apply the general discussion above to two specific physical systems, the Einstein--Maxwell theory and the Einstein--Euler--Heisenberg type of the effective field theory. 
The former case is an example of the 2-terms analysis and we find that no geodesic satisfies the sufficient conditions for time advance in the region where both $(G |M|)^{1/(D-3)} \ll r$ and $(G k Q^2)^{1/2(D-3)} \ll r$ are satisfied. 
Our results include the situation where the spacetime is super-extremal, or has a negative ADM mass. Hence, our results are beyond the situation included by the general discussion by Refs.~ \cite{Woolgar1994ar,Page:2002xn,Gao:2000ga}, where the positivity of the mass or the global hyperbolicity of the conformally completed spacetime is assumed.
The latter case is an application of the 3-terms case.
We find that if the parameters of the Euler--Heisenberg correction terms $\alpha_{1}$ and $\alpha_2$ satisfy Eq.~\eqref{timeadvancealpha}, that is, $0<\alpha_A<G\ell^2/(8(D-3)k^2)$, there always exists a choice of a solution with a positive ADM mass that admits time advance null geodesics, though we find that such a solution must be superextremal and the time advance null geodesics are passing through the region with the negative quasi--local energy. 
Our result indicates that if we require that the Einstein--Euler--Heisenberg type of the effective field theory prohibits time advance for any choice of the solution of the equations of motion, the parameters $\alpha_{1}$ and $\alpha_2$ with the conditions \eqref{ta condition:alpha1+2} and~\eqref{ta condition:alpha2} are excluded. Thus, the parameters $\alpha_{1}$ and $\alpha_2$ must satisfy
\begin{align}
&2\alpha_{1}+\alpha_2 \leq 0, \qquad \text{or}\qquad  \frac{\ell^2 G}{16 (D-3) k^2}\leq 2\alpha_{1}+\alpha_2,\label{constraintalpha1}\\
 &\alpha_{2} \leq 0, \ \ \ \ \ \ \ \ \  \ \ \ \;\; \;  \text{or}\qquad  \frac{\ell^2 G}{8 (D-3) k^2}\leq \alpha_{2}. \label{constraintalpha2}
\end{align}

In the flat limit $\ell \rightarrow \infty$, our constraints \eqref{constraintalpha1} and~\eqref{constraintalpha2} suggest that $2\alpha_1+\alpha_2$ and $\alpha_{2}$ must be negative to ensure the absence of time advance for {\it any} solution of the equations of motion of this system. 
On the other hand, it is known that the positivity of $2\alpha_1+\alpha_2$ and $\alpha_{2}$
\footnote{
For $D = 4$, our Euler--Heisenberg term can be expressed as 
\begin{align*}
 \frac{1}{4 \pi} \left( \alpha_{1} (F_{\mu\nu} F^{\mu\nu})^2 + \alpha_{2} F_{\mu\nu} F^{\nu\rho} F_{\rho\sigma} F^{\sigma\mu} \right) = \frac{2 \alpha_{1} + \alpha_{2}}{8 \pi} (F_{\mu\nu} F^{\mu\nu} )^2 + \frac{\alpha_{2}}{16 \pi} (F_{\mu\nu} \tilde{F}^{\mu\nu})^2.
\end{align*}
The positivity bound \cite{Adams:2006sv, Cheung:2014ega,Hamada:2018dde} requires the positivity of each coefficients in the right hand side.
}
 is required to maintain analyticity, unitarity, causality, and locality in the ultraviolet quantum field theory behind the Euler--Heisenberg effective field theory, at least in the absence of graviton exchange \cite{Adams:2006sv, Cheung:2014ega,Hamada:2018dde}\footnote{For more recent attempts to include the effect of graviton exchange, see, for example, Refs.~\cite{Bellazzini:2019xts, Tokuda:2020mlf}.}.
As examined at the end of Sec.~\ref{subsec:EH}, time advance occurs only when a geodesic passes through a region with negative quasi-local mass in a super-extremal spacetime 
and through a region where the null convergence condition for the effective metric is violated. Note that the null convergence condition for the physical metric remains satisfied in such regions. Therefore, the presence of such time advance geodesics would indicate a lack of physical reasonability of the solutions.

Finally, let us comment and outline the potential future directions.
Our conditions for the time advance, Eqs.~\eqref{def:calF} and~\eqref{def:calG}, derived in the general setup can be applied to any static, spherically symmetric, asymptotically AdS effective metric. Although we examined the application to the Einstein--Maxwell theory and the Einstein--Euler--Heisenberg effective field theory in this paper, our general formula has a wide range of applicability. For example, it is important including higher-curvature corrections and interactions between the electromagnetic field and gravity, such as 
$R_{\mu\nu\rho\sigma} F^{\mu\nu} F^{\rho\sigma}$, $R_{\mu\nu\rho\sigma} R^{\mu\nu\rho\sigma}$ terms, or even higher-order corrections like $R^3$
 \footnote{
The initial value formulation with such higher order derivative terms is discussed, for instance, in Ref. \cite{Davies:2021frz}.
}.
In addition, our current analysis is restricted to static, spherically symmetric cases. Therefore, it is interesting to generalize this framework to more complicated circumstances, such as, static-axially symmetric spacetimes, or dynamical spacetimes.

%
%
\begin{acknowledgments}
L.~F. is  supported by THERS Interdisciplinary Frontier Next Generation Researcher.
K.~I. and D.~Y. are supported by Grants-in-Aid for Scientific Research from 
the Ministry of Education, Culture, Sports, Science and Technology of Japan (MEXT)/ Japan Society for the Promotion of Science (JSPS), Grant Numbers JP21H05182 (K.~I.), JP21H05189 (K.~I. and D.~Y.), JP24K07046(K.~I.), and JP20K14469 (D.~Y.).
\end{acknowledgments}

\
\appendix

\section{Detailed Derivations}
\label{AppendixA}

In this appendix, we present the details of the derivations of the expression for $\Delta \phi$ and $\Delta t$, given by Eqs.~\eqref{timeadv1} and \eqref{timeadv2} respectively.
The derivation is given in appendix~\ref{app:deltaphit}. 
The form of $\Delta t$ involves the hypergeometric function \eqref{2F1integral}. 
In appendix~\ref{app:gamman}, some properties of the hypergeometric function are shown.

\subsection{Derivation of $\Delta \phi$ and $\Delta t$ }
\label{app:deltaphit}

A general static, spherically symmetric asymptotically AdS metric is given in Eq.~\eqref{general}. 
Since the conformal transformation keeps each null geodesics unchanged, 
we factor out a conformal factor such that the metric function of the spherical parts becomes simple,
\begin{align}
     \widehat{g}_{\mu\nu} dx^{\mu} dx^{\nu} 
= g(r) \tilde {g}_{\mu\nu} dx^{\mu} dx^{\nu} \coloneqq
g(r)
\left( - \tilde{f}(r) dt^2 + \frac{\tilde{h}(r)}{\tilde{f}(r)} dr^2 + r^2 d\Omega_{D-2}^2 \right),
\end{align}
where $\tilde{f}(r)$ and $\tilde{h}(r)$ are defined by 
\begin{align}
 \tilde{f}(r) &\coloneqq \frac{f(r)}{g(r)} = 1 + \frac{r^2}{\ell^2} + \sum_{n = 1}^{\infty} \frac{\tilde{f}_{n}}{r^n}+ {\cal O}\left(\epsilon^2 \right), \\
 \tilde{h}(r) &\coloneqq \frac{h(r)}{g(r)^2} = 1 + \sum_{n = 1}^{\infty} \frac{\tilde{h}_{n}}{r^n} + {\cal O}\left(\epsilon^2 \right),
\end{align}
with 
\begin{align}
 \tilde{f}_{n} &= f_{n} - g_{n} - \frac{g_{n+2}}{\ell^2}, \\
 \tilde{h}_{n} &= h_{n} - 2 g_{n}.
\end{align}
In the analysis of the null geodesic, we use $\tilde {g}_{\mu\nu}$ instead of $\widehat {g}_{\mu\nu}$, 
which makes the analysis simpler due to the absence of $\tilde {g}(r)$.

Let us evaluate $\Delta \phi$. By using $\tilde{f}(r)$ and $\tilde{h}(r)$, 
Eq.~\eqref{genep} can be expressed as 
\begin{align}
     &\Delta\phi=2\int^\infty_{r_m} d r \frac{b \sqrt{\tilde{h}(r)}}{r^2 \sqrt{1-\frac{b^2 \tilde{f}(r) }{r^2}}}.
\end{align}
By introducing a new variable $z = r_{m}/r$ and rewriting the impact parameter $b$ by $b = r_{m}/\sqrt{\tilde{f}(r_{m})}$, the integral can be expressed as 
\begin{align}
\Delta \phi &= 2 \int_{0}^{1} d z \frac{\sqrt{\tilde{h}(r_{m}/z)}}{\sqrt{\tilde{f}(r_{m}) - \tilde{f}(r_{m}/z) z^2}} 
= 2 I^{(1)}_{0} + \sum_{n = 1}^{\infty} \frac{C_{n}}{r_{m}^{n}} + \mathcal{O}(\epsilon^2), \label{deltaphibyIn}
\end{align}
where $C_{n}$ are defined by
\begin{align}
C_{n} \coloneqq  \tilde{h}_{n} I^{(1)}_{n} +  \tilde{f}_{n} I^{(2)}_{n}, \label{defCn}
\end{align}
and the integrals $I^{(1)}_{n}$ and $I^{(2)}_{n}$ are defined by
\begin{align}
I^{(1)}_{n} &\coloneqq \int_{0}^{1} dz ~ \frac{z^{n}}{\sqrt{1 - z^2}}, \\
 I^{(2)}_{n} &\coloneqq \int_{0}^{1} dz ~\left(- \frac{1}{(1 - z^2)^{3/2}} + \frac{z^{n+2}}{(1 - z^2)^{3/2}} \right).
\end{align}
Note that the $\ell$ dependence in the function $\tilde{f}$ is cancelled.
The integral $I^{(1)}_{n}$ can be expressed in terms of the integral representation of the beta function $B(p,q)$
\begin{align}
 B(p, q) &\coloneqq \frac{\Gamma(p) \Gamma(q)}{\Gamma(p+q)} = 2 \int_{0}^{1} d z~ z^{2p - 1} (1 - z^2)^{q - 1}, \qquad \left(p,q > 0\right),
\end{align}
as
\begin{align}
I^{(1)}_{n} &= \frac{1}{2} B\left(\frac{1}{2}, \frac{1+n}{2}\right).\label{I1}
\end{align}
Note that for $n = 0$, we obtain $I^{(1)}_{0} = \pi/2$. 
The integral $I^{(2)}_{n}$ can be expressed by $I^{(1)}_{n}$, and hence expressed by the beta function as well, through the following calculation,
\begin{align}
 I^{(2)}_{n}
 &= \int_{0}^{1} dz~ \left( - \frac{1}{(1 - z^2)^{\frac{3}{2}}} + z^{n + 1} \left( \frac{1}{\sqrt{1 - z^2}}\right)' \right) \notag\\
 &= \left[
- \frac{z}{\sqrt{1 - z^2}} + \frac{z^{n + 1}}{\sqrt{1 - z^2}}
\right]_{0}^{1} - (n + 1)I_{n}^{(1)} \notag\\
&= - \frac{n + 1}{2} B\left(\frac{1}{2}, \frac{1+n}{2}\right). \label{I2}
\end{align}
By substituting the expressions \eqref{I1} and \eqref{I2} into Eq.~\eqref{deltaphibyIn} with Eq.~\eqref{defCn}, we obtain
\begin{align}
\Delta \phi = \pi + \sum_{n = 1}^{\infty} \frac{C_{n}}{r_{m}} + \mathcal{O}(\epsilon^2),
\end{align}
with
\begin{align}
 C_{n} &=  \frac{1}{2} B\left(\frac{1}{2}, \frac{1+n}{2}\right) \left( \tilde{h}_{n} - (1+n) \tilde{f}_{n} \right) \\
&= \frac{1}{2} B\left(\frac{1}{2}, \frac{1+n}{2}\right) \left(
- (1 + n) f_{n} + h_{n} + (n - 1) g_{n} + \frac{1 + n}{\ell^2} g_{n+2}
\right).
\end{align}

Next, let us calculate $\Delta t/ \ell$ and derive Eq.~\eqref{timeadv2}.
By using the functions $\tilde{f}$ and $\tilde{h}$, the expression for $\Delta t/\ell$, Eq.~\eqref{genet}, can be written as 
\begin{align}
   &\Delta t = 2 \int^\infty_{r_m} d r \frac{\sqrt{\tilde{h}(r)}}{\tilde{f}(r)\sqrt{1-\frac{b^2 \tilde{f}(r)}{r^2}}}.
\end{align}
Again, by using the variable $z = r_{m}/r$, $\Delta t/\ell$ can be expressed as 
\begin{align}
 \frac{1}{\ell} \Delta t = 
&=
 2 J^{(1)}_{0} + \sum_{n=1}^{\infty} \frac{D_{n}}{r_{m}^{n}} + \mathcal{O}(\epsilon^2)\label{deltatbyJn}.
\end{align}
Here we define the coefficients $D_{n}$ by
\begin{align}
 D_{n} = \tilde{h}_{n} J^{(1)}_{n}  + \tilde{f}_{n} J^{(2)}_{n},\label{defDn}
\end{align}
and the integrals $J^{(1)}_{n}$ and $J^{(2)}_{n}$ by
\begin{align}
 J^{(1)}_{n}(w) &\coloneqq \int_{0}^{1} dz ~ \frac{z^{n}}{1 - w z^2} \sqrt{\frac{1 - w}{1 - z^2}}, \\ 
 J^{(2)}_{n}(w) & \coloneqq \int_{0}^{1} d z~ \left( - \frac{1}{\sqrt{1 - w}} \frac{1}{(1 - z^2)^{3/2}} + z^{n + 2} \frac{\sqrt{1 - w}}{(1 - z^2)^{\frac{3}{2}}} \frac{(1 + 2 w - 3 w z^2)}{(1 - w z^2)^2}  \right).
\end{align}
Also we introduce the variable $w < 0$ by 
\begin{align}
 w \coloneqq - \frac{\ell^2}{r_{m}^2}.
\end{align}
The integral $J^{(1)}_{n}(w)$ can be expressed by the integral representation for the hypergeometric function, 
\begin{align}
 {}_{2} F_{1} (a, b, c; w) = \frac{2}{B(a, - a + c)} \int_{0}^{1} dz~ z^{2 a - 1} ( 1 - z^2)^{c - a - 1} ( 1 - w z^2 )^{-b}, \label{2F1integral}
\end{align}
which holds for complex parameters $\{a,b,c\}$ satisfying $\text{Re } c > \text{Re } a > 0$.
Applying this expression with the parameters $a = \frac{1+n}{2}, b = 1$ and $c = \frac{n}{2} + 1$, one can express $J^{(1)}_{n}(w)$ as
\begin{align}
 J^{(1)}_{n}(w) = \frac{1}{2} B\left(\frac{n+1}{2}, \frac{1}{2} \right) \sqrt{1 - w} ~{}_{2}F_{1}\left(\frac{1 + n}{2},1, \frac{n}{2} + 1; w \right).
\end{align}
In addition, $w$ dependence can be simplified by using the Kummer's relation,
\begin{align}
{}_{2}F_{1}(a,b,c;w) = (1 - w)^{c - a - b} {}_{2}F_{1}(c - b, c - a, c; w).
\end{align}
The result is 
\begin{align}
  J^{(1)}_{n}(w) = \frac{1}{2} B\left(\frac{n+1}{2}, \frac{1}{2} \right) {}_{2}F_{1}\left(\frac{n}{2}, \frac{1}{2}, \frac{n}{2} + 1; w \right).\label{J1n}
\end{align}
Note that for $n = 0$, the integral can be evaluated as $J^{(1)}_{0}(w) = \pi/2$.
The integral $J^{(2)}_{n}$ can be expressed by $J^{(1)}_{n}$ through the integration by part, and hence expressed by the hypergeometric function, as follows:
\begin{align}
 J^{(2)}_{n}(w) & = \int_{0}^{1} d z~ \left( - \frac{1}{\sqrt{1 - w}}\frac{1}{(1 - z^2)^{3/2}} + z^{n + 1} \left( \frac{1}{1 - w z^2} \sqrt{ \frac{1 - w}{1 - z^2}}\right)' \right) \notag\\
&=
 \left[
- \frac{1}{\sqrt{1-w}} \frac{z}{\sqrt{1 - z^2}} + \frac{z^{n+1}}{1 -w z^2} \sqrt{\frac{1 - w}{1 - z^2}}
\right]^{1}_{0}  - (n + 1)J^{(1)}_{n}(w) \notag\\
&= - (n + 1) \frac{1}{2} B\left( \frac{n + 1}{2}, \frac{1}{2} \right) {}_{2}F_{1}\left( \frac{n}{2}, \frac{1}{2}, \frac{n}{2} + 1; w \right).\label{J2n}
\end{align}
Then, by substituting the expressions \eqref{J1n} and \eqref{J2n} into Eq.~\eqref{deltatbyJn} with the definition \eqref{defDn}, we obtain
\begin{align}
 \frac{1}{\ell} \Delta t = \pi + \sum_{n = 1}^{\infty} \frac{D_{n}}{r_{m}^{n}}+{\cal O}\left( \epsilon^2\right),
\end{align}
with
\begin{align}
 D_{n} &= \frac{1}{2} B\left( \frac{n + 1}{2}, \frac{1}{2} \right) {}_{2}F_{1}\left( \frac{n}{2}, \frac{1}{2}, \frac{n}{2} + 1; w \right) \left( \tilde{h}_{n} - (n + 1) \tilde{f}_{n} \right) \notag\\
& = {}_{2}F_{1}\left( \frac{n}{2}, \frac{1}{2}, \frac{n}{2} + 1; w \right) C_{n}.
\end{align}
Hence, we obtain
\begin{align}
 \frac{1}{\ell} \Delta t - \Delta \phi = - \sum_{n = 1}^{\infty} \beta_{n} \frac{C_{n}}{r_{m}^{n}}+{\cal O}\left( \epsilon^2\right),
\end{align}
with
\begin{align}
\beta_{n}(w) = 1 - {}_{2}F_{1}\left( \frac{n}{2}, \frac{1}{2}, \frac{n}{2} + 1; w \right).   
\end{align}

\subsection{Properties of $\beta_{n}(w)$}
\label{app:gamman}

In the analysis of time advance conditions, we need to understand the properties of $\beta_{n}(w)$. 
In this appendix, we investigate it. 
Note that $n$ and $w$ satisfy $n \geq 1$ and $w \leq 0$, respectively. 

From the integral expression of the hypergeometric function \eqref{2F1integral}, 
$\beta_{n}$ can be represented as
\begin{align}
 \beta_{n}(w) = 1 + n \int_{0}^{1} d z \frac{- z^{n-1}}{\sqrt{1 + |w| z^2}}.
\end{align}
Since the integrand is negative and an increasing function of $|w|$, we obtain $\beta_{n}(w) < 1$ and $\beta_{n}(w) > \beta_{n}(0) = 0$ respectively, and thus $0 < \beta_{n}(w) < 1$.
In addition, by the integration by part, $\beta_{n}(w)$ can be expressed as 
\begin{align}
 \beta_{n}(w) &= 1 - \int_{0}^{1} d z \frac{(z^{n})'}{\sqrt{1 + |w| z^2}} \notag\\
&= 1 - \left[ \frac{z^{n}}{\sqrt{1+|w| z^2}}\right]^{1}_{0} - |w| \int_{0}^{1} dz \frac{z^{n + 1}}{(1 + |w| z^2)^{\frac{3}{2}}} \notag\\
&= 1 - \frac{1}{\sqrt{1 + |w|}} - |w| \int_{0}^{1} dz \frac{z^{n + 1}}{(1 + |w| z^2)^{\frac{3}{2}}}.
\end{align}
This expression implies the monotonic increase of $\beta_{n}(w)$ with respect to $n$, {\it i.e.},
\begin{align}
 \beta_{n_{1}}(w) < \beta_{n_{2}}(w),
\end{align}
for any $1 \leq n_{1} < n_{2}$, because $- z^{n_{1}} < - z^{n_{2}}$ holds in $0 < z < 1 $.
Therefore, for a given $w < 0$, the lower and upper bounds for $\beta_{n}(w)$ are given by $\beta_{1}(w)$ and $\beta_{\infty}(w)$ respectively, which are explicitly written as   
\begin{align}
 \beta_{1}(w) &= 1 - \frac{\sinh^{-1} \sqrt{|w|}}{\sqrt{|w|}}, \\
 \beta_{\infty}(w) &= 1 - \frac{1}{\sqrt{1 + |w|}}.
\end{align}
Thus, we can strengthen the inequality $0 < \beta_{n}(w) < 1$ to
\begin{align}
 0 < 1 - \frac{\sinh^{-1} \sqrt{|w|}}{\sqrt{|w|}} < \beta_{n}(w) < 1 - \frac{1}{1 + |w|} < 1.
\end{align}
We also evaluate the ratio of $\beta_{n}$ as
\footnote{
The function $\left(1 - 1/\sqrt{1 + |w|}\right)/\left(1 - \sinh^{-1} \sqrt{|w|}/\sqrt{|w|}\right)$ is decreasing function of $|w|$ and the value in the limit $|w| \rightarrow 0$ is $3$.
}
\begin{align}
 1 < \frac{\beta_{n_{2}}(w)}{\beta_{n_{1}}(w)} < \frac{\beta_{\infty}(w)}{\beta_{1}(w)} = \frac{1 - \frac{1}{\sqrt{1 + |w|}}}{1 - \frac{\sinh^{-1} \sqrt{|w|}}{\sqrt{|w|}}} < 3, 
\end{align}
for $n_1<n_2$.
Thus, the ratio $\beta_{n_{2}}/ \beta_{n_{1}}$ can be regarded as $\mathcal{O}(1)$ quantity.

\subsection{Derivation of inequality~\eqref{b/b}}\label{secb/b}

Objective here is to prove the inequality~\eqref{b/b}, which is equivalent to proving
\begin{equation}
\frac{\beta_{2(D-2)}(\omega)}{\beta_{2(D-3)}(\omega)} < \frac{\beta_{2(D-2)}(0)}{\beta_{2(D-3)}(0)}
\end{equation}
This follows from the more general statement that for arbitrary integers $ n_2 > n_1 > 0 $,
\begin{equation}
\frac{\beta_{n_2}(w)}{\beta_{n_1}(w)} < \frac{\beta_{n_2}(0)}{\beta_{n_1}(0)}.
\label{bb<bb}
\end{equation}
Thus, it suffices to prove the latter inequality.
We show this inequality, relying partially on numerical results.

To show the inequality~\eqref{bb<bb}, it is sufficient to show the monotonic decrease of the ratio $\frac{\beta_{n_2}(w)}{\beta_{n_1}(w)}$ with respect to $\w$. Differentiating $\beta_n(w)$ with respect to $\w$, we obtain
\begin{equation}
\frac{d}{d\w} \beta_n(w) = \frac{n}{2} \int_0^1 \frac{z^{n+1}}{(1 + \w z^2)^{3/2}}\,dz.
\end{equation}
Introducing
\begin{equation}
I_n(\w) := \int_0^1 \frac{z^{n+1}}{(1 + \w z^2)^{3/2}}\,dz > 0,
\end{equation}
we rewrite
\begin{equation}
\beta_n(w) = 1 - \frac{1}{\sqrt{1 + \w}} - \w I_n(\w), \quad \frac{d}{d\w} \beta_n(w) = \frac{n}{2} I_n(\w).
\end{equation}
Differentiating the ratio yields
\begin{equation}
\frac{d}{d\w} \left( \frac{\beta_{n_2}(w)}{\beta_{n_1}(w)} \right) 
= \frac{\beta_{n_2}(w)}{\beta_{n_1}(w)} \left( \frac{\frac{d}{d\w} \beta_{n_2}(w)}{\beta_{n_2}(w)} - \frac{\frac{d}{d\w} \beta_{n_1}(w)}{\beta_{n_1}(w)} \right).
\end{equation}
Since the ratio $\frac{\beta_{n_2}(w)}{\beta_{n_1}(w)}$ is positive, it suffices to show that
\begin{equation}
\frac{\frac{d}{d\w} \beta_{n_2}(w)}{\beta_{n_2}(w)} - \frac{\frac{d}{d\w} \beta_{n_1}(w)}{\beta_{n_1}(w)} < 0,
\end{equation}
which reduces to proving that the function
\begin{equation}
\frac{\frac{d}{d\w} \beta_n(w)}{\beta_n(w)}
\end{equation}
is monotonically decreasing in $n$.
Equivalently, it suffices to prove the monotonic increase of the inverse function
\begin{equation}
\left(\frac{\frac{d}{d\w} \beta_n(w)}{\beta_n(w)}\right)^{-1}
= \frac{2}{n} \left[ \left( 1 - \frac{1}{\sqrt{1 + \w}} \right) \frac{1}{I_n(\w)} - \w \right],
\label{ivs}
\end{equation}
 with respect to $n$.

Taking the derivative of Eq.\eqref{ivs} with respect to $n$, we find
\begin{equation}
\frac{\partial}{\partial n} \left( \frac{\frac{d}{d\w} \beta_n(w)}{\beta_n(w)} \right)^{-1}
= -\frac{2}{n^2} \left[ \left( 1 - \frac{1}{\sqrt{1 + \w}} \right) \frac{I_n(\w) + n \frac{\partial}{\partial n} I_n(\w)}{I_n(\w)^2} - \w \right].
\end{equation}
Hence, proving positivity of the term in brackets completes the proof. We define
\begin{align}
F(n,\w) &:= \left(1 - \frac{1}{\sqrt{1 + \w}} \right) f(n,\w) - \w, \\
f(n,\w) &:= \frac{I_n(\w) + n \frac{\partial}{\partial n} I_n(\w)}{I_n(\w)^2},
\end{align}
and then, the goal reduces to showing $F(n,\w) < 0$.
This can be achieved by proving that $F(n,\w)$ is monotonically decreasing in $n$ with $F(0,\w) = 0$.

First, we verify that $F(0,\w)=0$. To this end, we note the following:
\begin{eqnarray}
I_0 (\w) :=\int^1_0 dz \frac{z}{(1+\w z^2)^{3/2}} =\frac{1}{\w} \left(1-\frac{1}{\sqrt{1+\w}} \right)
\end{eqnarray}
and
\begin{eqnarray}
\left| \dn I_n (\w)\big|_{n=0} \right| = \left| \int^1_0 dz \frac{ (\ln z) z}{(1+\w z^2)^{3/2}}\right|
\le \int^1_0 dz |\ln z| z < \infty.
\end{eqnarray}
From these, it follows that
\begin{eqnarray}
f(0,\w)=\frac{I_0 (\w) + 0 \cdot \dn I_n (\w) \big| _{n=0}}{I_0^2(\w)} = \frac{1}{I_0(\w)}.
\end{eqnarray}
Therefore,
we have
\begin{eqnarray}
F(0,\w)= \left(1-\frac{1}{\sqrt{1+\w}} \right) f(0,\w) -\w = \left(1-\frac{1}{\sqrt{1+\w}} \right) \frac{1}{I_0(\w)} -\w = 0. \nn
\end{eqnarray}

\begin{figure}[t]
\centering
\begin{minipage}[b]{0.49\columnwidth}
    \centering
    \includegraphics[width=0.9\columnwidth]{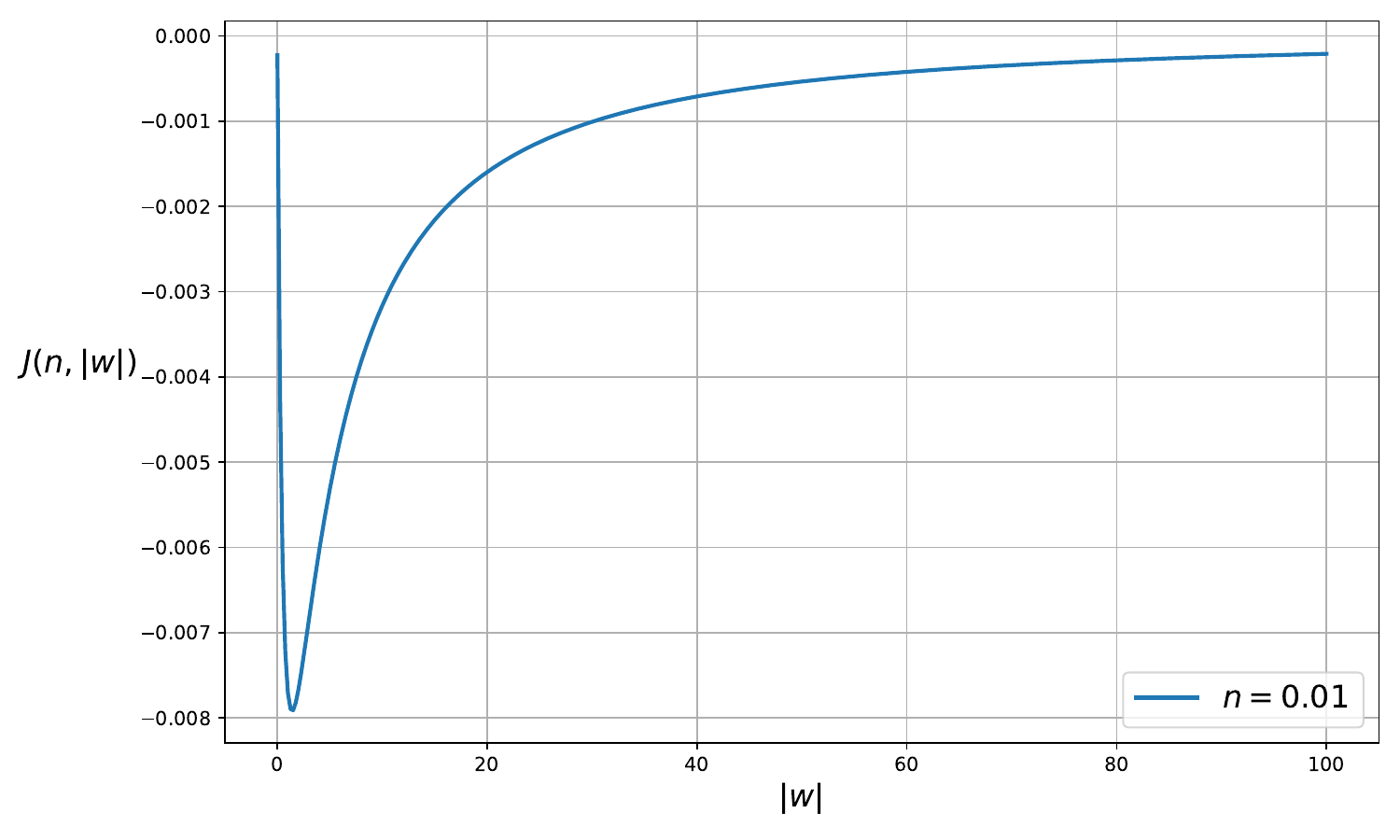}
    \vfill
    \includegraphics[width=0.9\columnwidth]{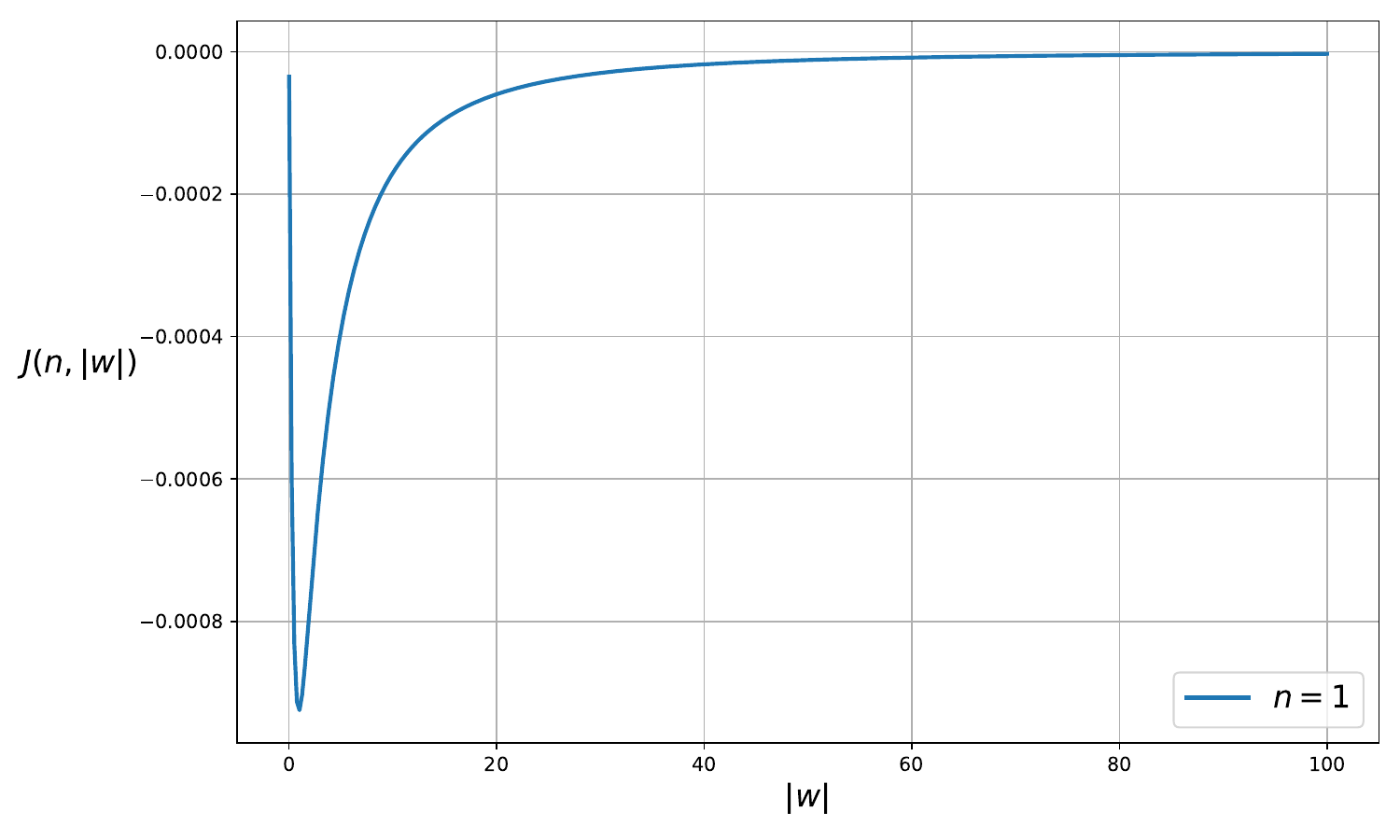}
    \vfill
    \includegraphics[width=0.9\columnwidth]{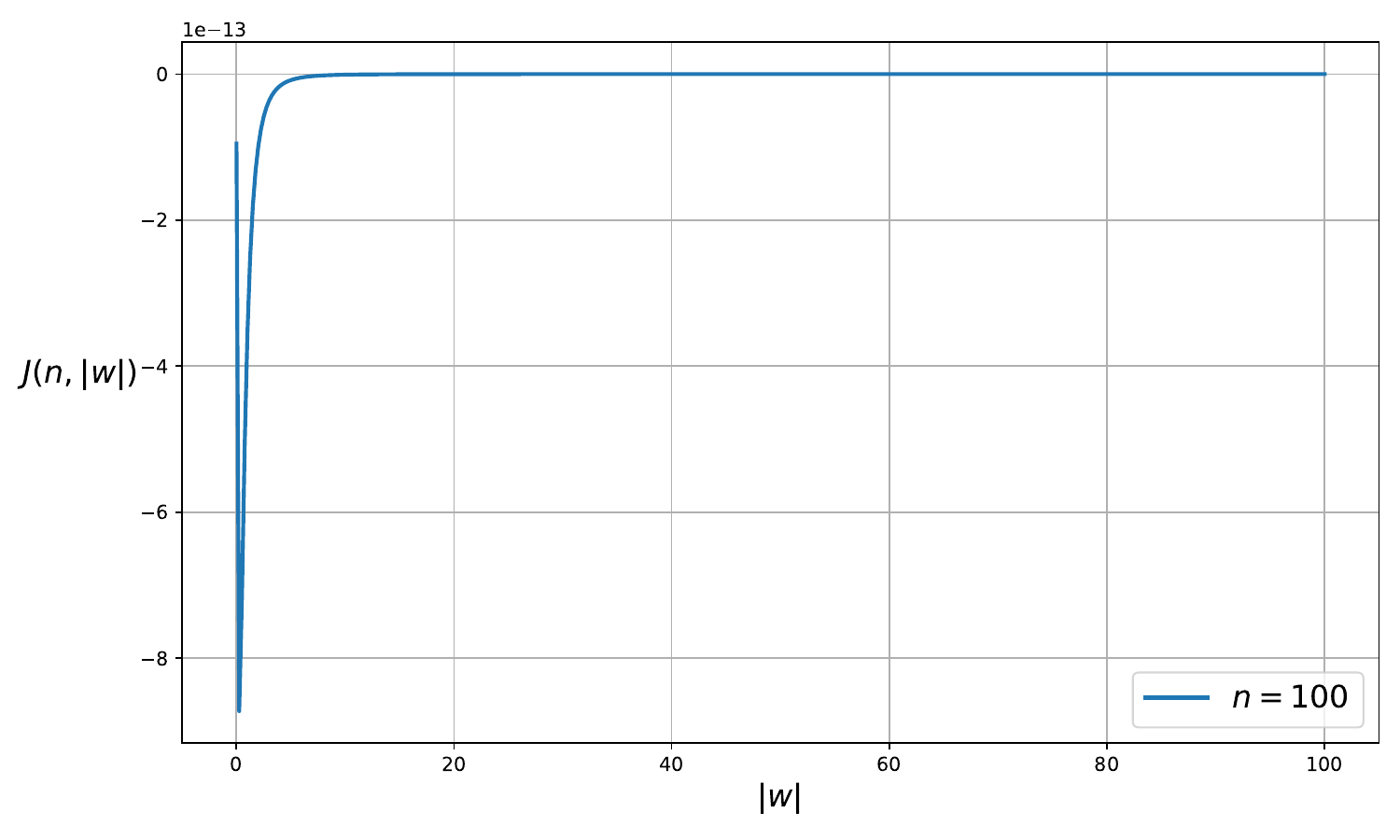}
\end{minipage}
\begin{minipage}[b]{0.49\columnwidth}
    \centering
    \includegraphics[width=0.9\columnwidth]{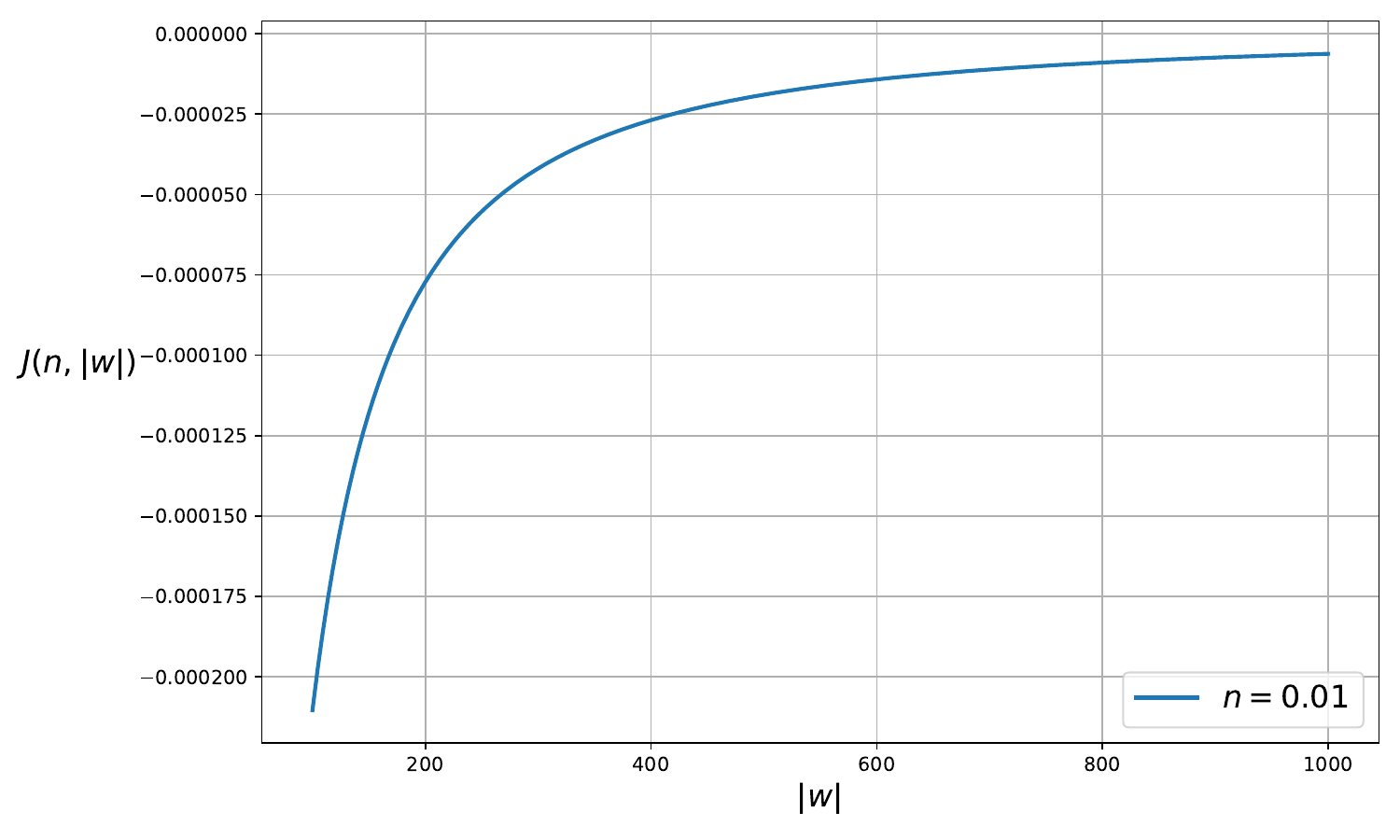}
    \vfill
    \includegraphics[width=0.9\columnwidth]{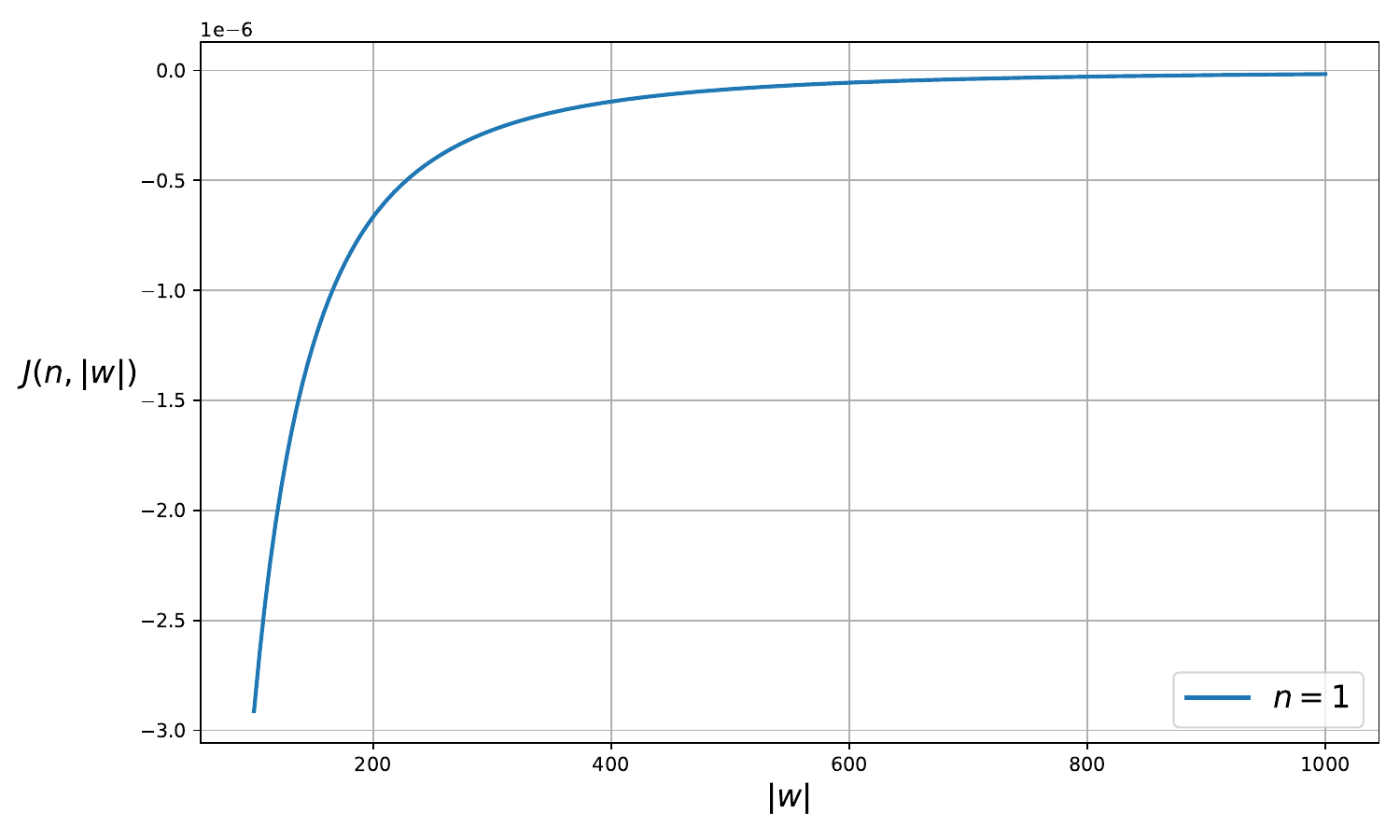}
    \vfill
    \includegraphics[width=0.9\columnwidth]{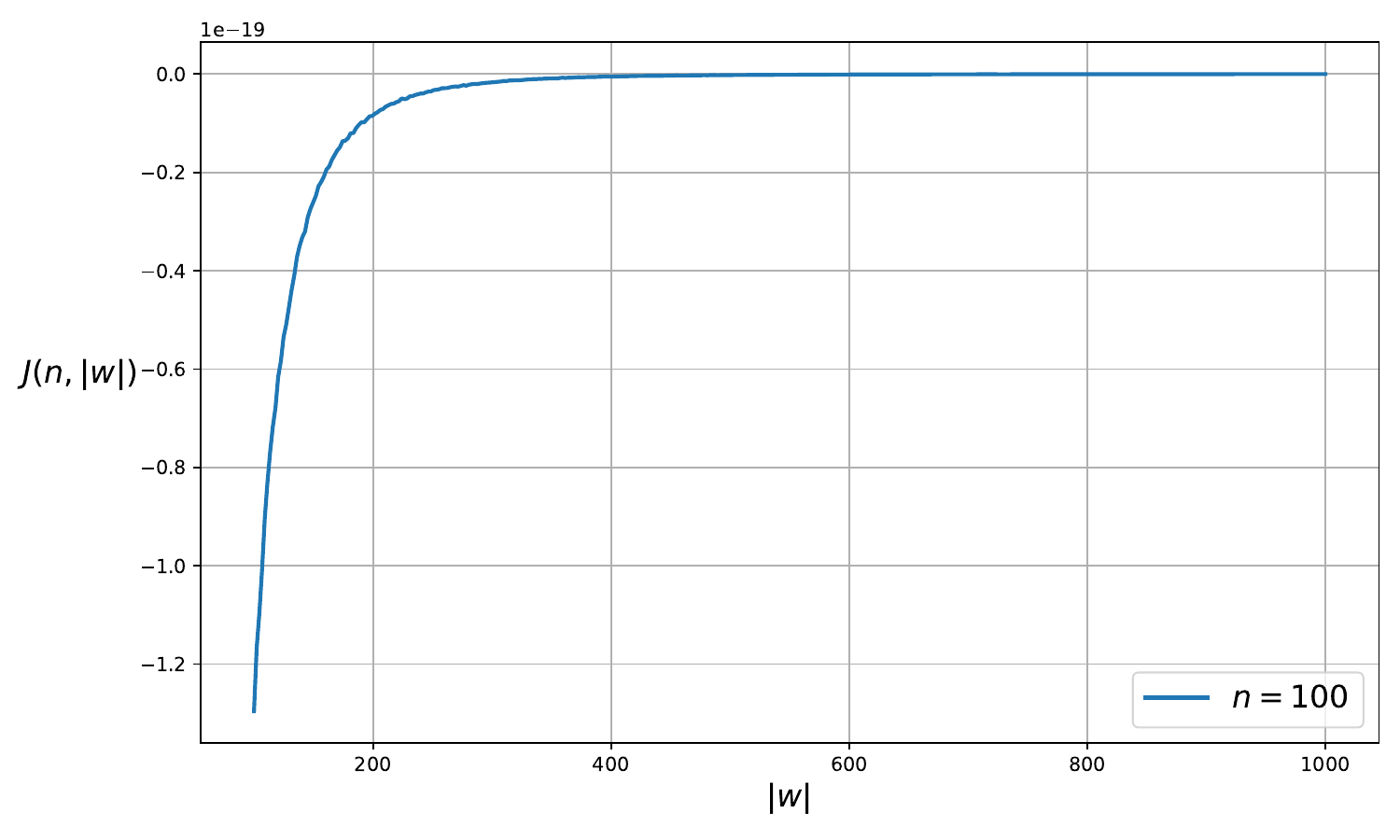}
\end{minipage}
    \caption{Plots of $J(n,\w)$ as a function of $\w$ with fixed $n$. The first, second, and third rows correspond to the case with $n=0.01$, $1$, $100$, respectively. The left and right columns show the plots for $0<\w<100$ and $100<\w<1000$, respectively.}
    \label{6graph}
\end{figure}

Next, we demonstrate the monotonic decrease of $F(n,\w)$ with respect to $n$.
The $n$-dependence of $F(n,\w)$ appears solely through the function $f(n,\w)$. 
Therefore, to show that $F(n,\w)$ is monotonically decreasing in $n$, it suffices to prove that $f(n,\w)$ is monotonically decreasing in $n$, 
since the prefactor $1 - \frac{1}{\sqrt{1+\w}}$ is strictly positive.
The derivative of $f(n,\w)$ with respect to $n$ is given by
\begin{eqnarray}
\dn f(n,\w)=\frac{n}{I_n^3(\w)}
\left[
I_n (\w) \left\{\left( \dn \right)^2 I_n (\w)\right\}-2\left\{ \dn I_n(\w)\right\}^2
\right].
\end{eqnarray}
Thus, it is sufficient to show that
\begin{eqnarray}
J(n,|w|):=I_n (\w) \left\{\left( \dn \right)^2 I_n (\w)\right\}-2\left\{ \dn I_n(\w)\right\}^2
<0, \qquad (\mbox{for}\ n>0)
\label{II-II}
\end{eqnarray}
in order to establish the desired monotonicity.
The numerical values of $J(n,\w)$ are shown in Fig.~\ref{6graph}, 
which indicates that $J(n,\w)$ is always negative.

\section{Time advance analysis of the case $\pi<\Delta\phi<2\pi$}
\label{app:largephi}
In the main section, we focus only on the case $0 < \Delta \phi \leq \pi$. In this appendix, we investigate the case with $\pi<\Delta\phi<2\pi$ and show that the conditions on the metric functions for the time advance are the same as in the case with $0\leq\Delta\phi\leq\pi$.

For $\pi<\Delta\phi<2\pi$, the coordinate of the end point $q$ of the boundary--to--boundary null geodesic (see Sec.~\ref{sec:bb causality} for the details) is 
\begin{align}
\phi' = -(2\pi-\Delta \phi) \quad(<0). 
\end{align}
Then, the time advance conditions, that is, the violation of Eq.~\eqref{t>|p|} is written as 
\begin{align}
    &\begin{cases}
        \pi<\Delta\phi<2\pi, \\
        \Delta t - \ell(2\pi - \Delta\phi) < 0.
    \end{cases} 
\end{align}
In the general static, spherically symmetric asymptotically AdS metric \eqref{general}, $\Delta t - \ell(2\pi - \Delta\phi)$ is written as
\begin{align}
     \Delta t/\ell-(2\pi-\Delta\phi)
     &=\Delta t/\ell-\Delta\phi+2(\Delta\phi-\pi)\nonumber\\
     &= \sum_{n=1}^{\infty} (2-\beta_{n}) \frac{C_{n}}{r_m^{n}} + \mathcal{O}(\epsilon^2).
\end{align}
Therefore, the time advance conditions are written in
\begin{align}
    &0<\sum_{n=1}^\infty \frac{C_n}{r_m^n}<\pi,\label{largephi condition1}\\
    &\sum_{n=1}^\infty(2-\beta_n) \frac{C_n}{r_m^n}<0.\label{largephi condition2}
\end{align}
Since the condition $\epsilon\ll1$ can be expressed as $|C_n|/r_m^n\ll1$, $\sum_{n=1}^\infty C_n/r_m^n$ cannot attain a value close to $\pi$, and thus, Eq.~\eqref{largephi condition1} can be simply written as 
\begin{align}
    \sum_{n=1}^\infty \frac{C_n}{r_m^n}>0.\label{largephi condition3}
\end{align}
After performing the transformation
\begin{align}
    \beta_n'\coloneqq\frac{1}{2-\beta_n},\ \ \ \ C_n'\coloneqq(2-\beta_n)C_n,
\end{align}
Eqs.~\eqref{largephi condition2} and~\eqref{largephi condition3} can be rewritten as 
\begin{align}
    \mathcal{F}'&\coloneqq\sum_{n=1}^\infty \frac{C_n'}{r_m^n}<0,\label{timeav4}\\
    \mathcal{G}'&\coloneqq\sum_{n=1}^\infty \beta_n'\frac{C'_n}{r_m^n}>0.\label{timeav5}
\end{align}
With the properties of $\beta_n$ given in Eqs.~\eqref{gamma1} and~\eqref{gamma2}, we can derive the corresponding properties of $\beta_n'$ as follows:
\begin{align}
    &\frac{1}{2}<\beta_n'<1,\label{gamma'1}\\
    &1<\frac{\beta'_{n_2}}{\beta_{n_1}'}  < \frac{\beta'_{\infty}}{\beta_{1}'} < \frac{1}{\,\, 1/2\,\, } =2 , \ \ \text{for}\ \  n_2>n_1.\label{gamma'2}
\end{align}
Therefore, by replacing $\mathcal{F}$, $\mathcal{G}$, $C_n$ and $\beta_n$ with $\mathcal{F}'$, $\mathcal{G}'$, $C_n'$ and $\beta_n'$ respectively, 
the analysis conducted in Sec.~\ref{sec:ta condition} applies in parallel. 

Since $(2-\beta_n')$ is of order unity, the order of $C_n'$ is the same as that of $C_n$. 
Then, the conditions for the perturbative expansion written in $C_n$ directly apply to this case. 
Thus,
supposing Eq.~\eqref{2terms1} holds for the 2-terms case, and Eqs.~\eqref{en2en3} and~\eqref{3term3} are satisfied for the 3-terms case, time-advance null geodesics exist in each case.
As a result, the time advance conditions are the same as those in the case with $0\leq\Delta\phi\leq\pi$.

\section{Einstein--Euler--Heisenberg theory and Effective Metrics}
In this appendix, we present the detailed analysis about the Einstein--Maxwell theory with the higher derivative corrections. 
The action is given in Eq.~\eqref{Action EH}.
A goal of this appendix is to derive the effective metrics for the electromagnetic wave.

\subsection{Equations of Motion}
The modified Einstein equation derived from the action~\eqref{Action EH} is written as
\begin{align}
\frac{1}{(D-2) G } \left( R_{\mu\nu}-\frac{1}{2}Rg_{\mu\nu}+\Lambda g_{\mu\nu} \right) -  T_{\mu\nu} = 0 \label{g2},
\end{align}
where $T_{\mu\nu}$ is the energy momentum tensor of electromagnetic field with Euler--Heisenberg correction terms given by 
\begin{align}
T_{\mu\nu} &\coloneqq \frac{1}{k} \left( F_{\mu\alpha}F_{\nu}^{\ \alpha}-\frac{1}{4}F^2 g_{\mu\nu} \right)
\notag\\
&\qquad +\alpha_1\left( (F_{\rho\sigma} F^{\rho\sigma})^2 g_{\mu\nu}- 8 F_{\rho\sigma}F^{\rho\sigma} F_{\mu}^{\ \beta}F_{\nu\beta}
	\right) \notag\\
&\qquad +\alpha_2\left(
	F^{\alpha\beta}F_{\rho\beta}F^{\sigma\rho}F_{\sigma\alpha}g_{\mu\nu} - 8 F_{\mu}^{\ \lambda}F_{\nu}^{\ \delta}F_{\lambda}^{\ \rho}F_{\delta\rho}
	\right).
\label{l2}
\end{align}
Similarly, the modified Maxwell equation derived by the variation of the action \eqref{Action EH} with respect to the Maxwell field $A_\mu$, is written as
\begin{align}
\frac{1}{k}\nabla_\nu F^{\mu\nu}-S^{\mu} = 0 \label{modifiedmaxwell},
\end{align}
where
\begin{align}
	S^{\mu}&\coloneqq8\alpha_1 (2F_{\rho\sigma}\nabla_\nu F^{\rho\sigma}F^{\mu\nu}+F^2\nabla_\nu F^{\mu\nu})
	\nonumber\\
	&\ 
	\ \ \ +8\alpha_2(
	\nabla_\nu F_{\sigma\rho}F^{\sigma\nu}F^{\mu\rho}+F_{\sigma\rho}\nabla_\nu F^{\sigma\nu}F^{\mu\rho}+F_{\sigma\rho}F^{\sigma\nu}\nabla_{\nu}F^{\mu\rho}).
\end{align}

\subsection{Static Spherically Symmetric Solutions}
\label{App:bgsolution}
Let us derive the static, spherically symmetric solution of the modified Einstein--Maxwell equations, considering the linear perturbations around the Reissner--Nordstrom--anti de Sitter solution with respect to the coupling constants $\alpha_{i}$.
Such treatments are valid when the corrections with $\alpha_{i}$ are not the leading order contributions in the equations of motion. 
This is satisfied if the terms with $\alpha_{i}$ are much smaller than (at least) one of the terms in the Einstein--Maxwell Lagrangian.
The leading-order contribution of the Einstein--Maxwell Lagrangian is the order of 
\begin{align}
\frac{|\Lambda|}{G}\sim \frac{1}{G \ell^2}, \qquad \mbox{or} \qquad \frac{1}{k} F^2 = \frac{k Q^2}{r^{2(D-2)}} \sim \frac{\epsilon}{G r^2}.
\end{align}
Here, the parameter $\epsilon$ represents $\epsilon \sim G M/r^{D-3}  ,G k Q^2/r^{2(D-3)} \ll 1$, which we assumed in the analysis of the time advance (see above Eq.~\eqref{timeadv1}).
 Since the terms with $\alpha_{i}$ are expressed as 
\begin{align}
  \alpha_{i} F^4 = \alpha_{i} \frac{ k^4 Q^4}{r^{4(D - 2)}} \sim \alpha_{i} k^2 \left( \frac{\epsilon}{G r^2} \right)^2, 
\end{align}
for the smallness of the $\alpha_{i}$ terms, either  
\begin{align}
\varepsilon_{\alpha}& \sim  \frac{k^2 |\alpha_{i}|}{G r^2} \epsilon \ll 1
\quad \mbox{or}\quad 
\frac{\ell^2}{r^2} \frac{k^2 |\alpha_{i}|}{G r^2} \epsilon^2 \ll 1 \label{validity123} 
\end{align}
is required to be satisfied.

Under these assumptions, the spacetime metric 
\begin{align}
 \bar{g}_{\mu\nu} dx^{\mu} dx^{\nu} = - \bar{f}(r) dt^2 + \frac{\bar{h}(r)}{\bar{f}(r)} dr^2 + r^2 \bar{g}(r) d \Omega^2_{D-2},
\end{align}
and the electrostatic potential
\begin{align}
 \bar{A}_{\mu} dx^{\mu} = - \bar{\Phi}(r) dt,
\end{align}
 are obtained as 
\begin{align}
 \bar{f}(r) &= 1 + \frac{r^2}{\ell^2} - \frac{2 G M}{r^{D-3}} + \frac{1}{D-3}\frac{G k Q^2}{r^{2(D-3)}} \notag\\
& \qquad - 
\frac{4}{3 D - 7} \frac{k^2 (2 \alpha_{1} + \alpha_{2})}{G r^2} \left( \frac{G k Q^2}{r^{2(D-3)}}\right)^2
+ \mathcal{O}
\left( \varepsilon_{\alpha}^2 \right) \notag\\
& = 1 + \frac{r^2}{\ell^2} - \frac{2 G M}{r^{D-3}} + \frac{1}{D-3}\frac{G k Q^2}{r^{2(D-3)}}
+ \mathcal{O}
\left( \epsilon \varepsilon_{\alpha}, \varepsilon_{\alpha}^2 \right)
,\label{fbarEH} \\ 
 \bar{h}(r) &=  1, \\
 \bar{g}(r) &= 1, \label{gbarEH}
\end{align}
and 
\begin{align}
 \bar{\Phi}(r)  &= \frac{k}{D-3} \frac{Q}{r^{D-3}} \left( 1  - 
\frac{8 (D-3)}{3 D - 7} \frac{k^2 (2 \alpha_{1} + \alpha_{2})}{G r^2} \frac{G k Q^2}{r^{2(D-3)}}
+
\mathcal{O}\left(
\varepsilon_{\alpha}^2
\right)
\right)
.
\end{align}
These results are obtained in a manner analogous to studies in the context of nonlinear electrodynamics~\cite{Yajima:2000kw,Ruffini:2013hia,Magos:2020ykt,Amaro:2020xro,Breton:2021mju,Nomura:2021efi,Abe:2023anf,Zhao:2024phz} and effective field theory~\cite{Kats:2006xp, Hamada:2018dde, Cheung:2018cwt, Jones:2019nev, Chen:2020hjm, Izumi:2024rge, Lin:2022ndf}, where the analysis has been limited to the cases with $D = 4$, without the cosmological constant, or both. 
As long as we focus on the leading order contributions of order $\mathcal{O}(\epsilon, \varepsilon_{\alpha})$, the contribution from the order $\mathcal{O}(\epsilon \varepsilon_{\alpha})$ can be neglected. 
Therefore, the expression \eqref{fbarEH} indicates that we can use Reissner--Nordstrom solution as the background metric even in the analysis of the Einstein--Euler--Heisenberg theory.

\subsection{Effective Metrics for Photon Propagations}
\label{app:EFT}
In the Einstein--Euler--Heisenberg theory, the kinetic terms of the Maxwell field are not in the canonical form. 
Therefore, the orbits of the fastest propagation are not described by the null geodesics with respect to the spacetime metric. 
Causality can be understood through the characteristics~\cite{courant1989methods}, 
and is often expressed in terms of the effective metric. 
We present the characteristics of the Einstein--Euler--Heisenberg theory in appendix~\ref{app:characteristic}, 
and subsequently derive the effective metrics for the scalar and the vector modes of the Maxwell field in appendix~\ref{App:emp}. 
Note that the effective metrics for gravitons are the same as those in the Einstein--Maxwell theory, that is, they match the spacetime metric.

\subsubsection{Characteristic Matrix}
\label{app:characteristic}
Here we investigate the characteristics of the modified Einstein--Maxwell equation in the Einstein--Euler--Heisenberg theory.
Let us consider the first-order perturbations around the static, spherically symmetric solution presented above,
\begin{align}
 g_{\mu\nu} &= \bar{g}_{\mu\nu} + h_{\mu\nu}, \\
 A_{\mu} &= \bar{A}_{\mu} + \delta A_{\mu},
\end{align}
and read the structures of the kinetic terms.

Since the highest-order derivative terms of the perturbations are essential for the causal structure, as discussed in the context of characteristic hypersurfaces~\cite{courant1989methods}, we focus only on these terms. 
The equation for the characteristics is obtained by replacing the partial derivative $\partial_\mu$ with the normal vector $\zeta_\mu$ of the characteristic surface
in the highest-order derivative terms. 
We adopt the notation commonly used in the analysis of characteristics, specifically using the symbol $\dot{=}$ to denote the operation where only the terms containing the highest-order derivatives are shown, specifically those with second-order derivatives in our case.

Since the second order derivatives are included only by the following quantities, 
\begin{align}
 \nabla_{\mu} F_{\nu\rho}& ~\dot{=}~ \left(\zeta_{\mu} \zeta_{\nu} \delta A_{\rho} - \zeta_{\mu} \zeta_{\rho} \delta A_{\nu} \right),
\\ 
R_{\mu\nu}&~\dot{=}~\frac{1}{2}\left(
	2\zeta_\alpha\zeta_{(\mu} h_{\nu)}^{\ \alpha}
-\zeta^2h_{\mu\nu}-\zeta_\mu\zeta_\nu h	\right),\\
R&~\dot{=}~\zeta_\mu\zeta_\nu h^{\mu\nu}-\zeta^2 h,
\end{align}
the highest derivative terms in the modified Einstein--Maxwell equations~\eqref{g2} and~\eqref{modifiedmaxwell} can be expressed as 
\begin{align}
&\frac{1}{(D-2) G} (R_{\mu\nu}-\frac{1}{2}Rg_{\mu\nu}+\Lambda g_{\mu\nu})- T_{\mu\nu}
~\dot{=}~
\frac{1}{(D-2) G} \mathcal{P}_{\mu\nu}{}^{\rho\sigma} h_{\rho\sigma},
\end{align}
and
\begin{align}
&\frac{1}{k} \nabla_\nu F^{\mu\nu}-S^{\mu} ~\dot{=}~ \frac{1}{k} \mathcal{P}^{\mu; \rho} \delta A_{\rho},
\end{align}
where
$\mathcal{P}_{\mu\nu;\rho\sigma}$ and 
$\mathcal{P}^{\mu; \rho}$ are given by 
\begin{align}
 \mathcal{P}_{\mu\nu;\rho\sigma}
&=  
  \zeta_\sigma \zeta_{(\mu}\bar{g}_{\nu)\rho}
- \frac{1}{2} \zeta^2 \bar{g}_{\mu(\rho} \bar{g}_{\sigma)\nu}
- \frac{1}{2} \zeta_\mu \zeta_\nu \bar{g}_{\rho\sigma}
- \frac{1}{2} \zeta_\rho \zeta_\sigma \bar{g}_{\mu\nu}
+\frac{1}{2} \bar{g}_{\mu\nu} \bar{g}_{\rho\sigma}\zeta^2,
\label{Gcmat}
\end{align} 
and
\begin{align}
\mathcal{P}^{\mu; \rho}
& = 
 \left(
    \zeta^\mu\zeta^\rho-\zeta^2 \bar{g}^{\mu\rho}
\right)
+8 k \alpha_1\left(
    \bar{F}^2 \zeta^2 \bar{g}^{\mu\rho} - \bar{F}^2 \zeta^\mu \zeta^\rho + 4\bar{F}^{\mu\nu} \bar{F}^{\rho\tau} \zeta_\nu \zeta_\tau
    \right)
    \nonumber \\ & \ 
    +8 k \alpha_2\big(
    \bar{F}^{\mu\nu} \bar{F}^{\rho\sigma}\zeta_\nu\zeta_\sigma + \bar{F}^{\mu\tau} \bar{F}^{\rho}_{\ \ \tau}\zeta^2 + \bar{F}_{\tau}{}^{\nu} \bar{F}^{\tau\beta}\zeta_\nu\zeta_\beta \bar{g}^{\mu\rho} - \bar{F}_{\sigma\tau} \bar{F}^{\mu\tau}\zeta^\rho\zeta^\sigma - \bar{F}_{\sigma\tau} \bar{F}^{\rho\tau}\zeta^\mu\zeta^\sigma
    \big). \label{EMcmat}
\end{align}
Since the characteristic matrix for both gravitational waves and electromagnetic waves is block-diagonal, we can discuss each wave separately. 
The characteristic matrix for gravitational waves~\eqref{Gcmat} is the same as that in the Einstein--Maxwell theory, 
and thus the causality for gravitational waves is expressed by the spacetime metric as usual. Therefore, we will focus only on analyzing the characteristic matrix for the electromagnetic waves.

\subsubsection{Effective Metric for Photons}
\label{App:emp}
Let us derive the effective metric for photons. Subscripts and superscripts $i$, $j$ represent the coordinates of $S_{D-2}$ in Eq.~\eqref{general}, while
subscripts and superscripts $a$, $b$ correspond to those of the other two coordinates in $\overline{g}_{ab}dx^adx^b=-fdt^2+(h/f)dr^2$. 
Let $D_i$ denote the covariant derivative with respect to the metric on $S_{D-2}$.
Due to the symmetry of $S_{D-2}$,  $\delta A_\mu$ can be decomposed into scalar and vector modes of $S_{D-2}$.
The components of $\delta A_a$ behave as scalars, while $\delta A_i$ includes one scalar and several  vector modes. 
This decomposition utilizes the proposition that outlines the decomposition of vectors and symmetric tensors on compact manifolds, as demonstrated in Ref.~\cite{Ishibashi:2004wx}. 
The decomposition of $\delta A_{\mu}$ is given by
\begin{align}
    &\delta A^a=\delta A^{(0)a},\label{exam3}\\
    &\delta A^i=D^i \delta A^{(0)}+\delta A^{(1)i},\label{exam3-2}
\end{align}
with
\begin{align}
    &D_i \delta A^{(1)i}=0,\label{exam4}
\end{align}
where the labels $^{(0)}$ and $^{(1)}$ denote the scalar and vector components, respectively.

\paragraph{Vector mode}
First, we focus on the vector mode $\delta A^{(1)i}$. 
The vector components satisfy Eq.~\eqref{exam4}, which means that, in terms of $\zeta^\mu$ the basis $e^{(1,I)\mu}$ of vector mode satisfies
\begin{align}
    &\zeta^\mu e^{(1,I)}_\mu=0,
\end{align}
where $I$ is for the label of the orthonormal basis of vector modes, that is, $g_{\mu\nu} e^{(1,I)\mu} e^{(1,J)\nu}= \delta^{IJ}$. 
Since the $t$- and $r$-components of the vector modes vanish, any vector mode is composed of $(D-3)$ basis vectors $e^{(1,I)}_\mu$. 
Operating the vector bases  $e^{(1,I)}_\mu$ and $e^{(1,J)}_\rho$ on the characteristic equation~\eqref{EMcmat}, we obtain 
\begin{align}
\mathcal{P}^{\mu;\rho} e^{(1,I)}_\mu e^{(1,J)}_\rho= 
	&\Bigg(\bar{g}^{\mu\nu} -8 k \alpha_1 \bar{F}^2 \bar{g}^{\mu\nu} -8 k \alpha_2 \bar{F}^{\mu \rho} \bar{F}^{\nu}{}_{\rho}  \Bigg) \zeta_\mu \zeta_\nu  \delta^{IJ} .
\end{align}
Then, the characteristic surface for every vector mode is given by 
\begin{align}
 \Bigg(\bar{g}^{\mu\nu} -8 k \alpha_1 \bar{F}^2 \bar{g}^{\mu\nu} -8 k \alpha_2 \bar{F}^{\mu \rho} \bar{F}^{\nu}{}_{\rho}  \Bigg) \zeta_\mu \zeta_\nu = 0.
\end{align}
By applying this equation recursively, one can see that the $\alpha_{1}$ term is higher order in $\varepsilon_{\alpha}$. Thus, we obtain
\begin{align} 
 \Bigg(\bar{g}^{\mu\nu} - 8 k \alpha_2 \bar{F}^{\mu \rho} \bar{F}^{\nu}{}_{\rho}  \Bigg) \zeta_\mu \zeta_\nu = 0,
\end{align}
and we can read the inverse of the effective metric as
\begin{align}
 (\widehat{g}{~}^{-1})^{\mu\nu} = \bar{g}^{\mu\nu} - 8 k \alpha_2 \bar{F}^{\mu \rho} \bar{F}^{\nu}{}_{\rho} .
\end{align}

\paragraph{Scalar mode}
Now, we derive the effective metric of the scalar mode.
The scalar degrees of freedom of $\delta A^\mu$ are $\delta A^{(0)a}$ and $\delta A^{(0)}$
in Eqs.~\eqref{exam3} and \eqref{exam3-2}, the number of which is three. 
Since the theory is invariant under the $U(1)$-gauge transformation $A_\mu \to A_\mu+ \partial_\mu \Lambda$, $A_\mu$ includes the gauge degree of freedom, 
which appears in the scalar sector.
In the analysis of the characteristic matrix, the components corresponding to the gauge degrees of freedom should be removed. 
In this paper, the directions of the propagations that we consider have the angular directions\footnote{Even if we consider the propagation without the angular components, one can obtain the same result. To avoid specifying every possible case, we are assuming this here.}, 
and thus, $t$- and $r$-components and $\zeta^\mu$ can be the basis of scalar mode. 
Since $\zeta^\mu$ corresponds to the gauge mode, the characteristic matrix should only be constructed using the $t$- and $r$-components.

Using the fact that in our analysis $\bar{F}^{\mu\nu}$ has only the $(t,r)$-component,  
the $(t,t)$-, $(r,r)$- and $(t,r)$-components of Eq.~\eqref{EMcmat} are calculated as
\begin{align}
&\mathcal{P}^{t; t}= \left(-\left[1-4k(2\alpha_1+\alpha_2)\bar{F}^2 \right] (\zeta^2-\zeta^t \zeta_t) 
+8k(2\alpha_1+\alpha_2)\bar{F}^2 \zeta^r \zeta_r\right)\bar{g}^{tt}, \\
&\mathcal{P}^{r; r}=\left(-\left[1-4k(2\alpha_1+\alpha_2)\bar{F}^2 \right] (\zeta^2-\zeta^r \zeta_r) 
+8k(2\alpha_1+\alpha_2)\bar{F}^2 \zeta^t \zeta_t\right)\bar{g}^{rr}, \\
&\mathcal{P}^{t; r}=\mathcal{P}^{r; t} =
\left[1-12k(2\alpha_1+\alpha_2)\bar{F}^2 \right]\zeta^t\zeta^r.
\end{align} 
The determinant of the characteristic matrix
\begin{align}
\mathcal{M}_s \coloneqq
\begin{pmatrix}
   \mathcal{P}^{t; t} & \mathcal{P}^{t; r} \\
   \mathcal{P}^{r; t} & \mathcal{P}^{r; r}
\end{pmatrix}
\end{align} 
for the scalar mode is 
\begin{align}
\det \mathcal{M}_s &= \left[1-4k(2\alpha_1+\alpha_2)\bar{F}^2 \right]^2
\zeta^i \zeta_i \bar{g}^{tt} \bar{g}^{rr} \left[\zeta^2 - \frac{8k(2\alpha_1+\alpha_2)\bar{F}^2}{1-4k(2\alpha_1+\alpha_2)\bar{F}^2}  \left(\zeta^t\zeta_t+\zeta^r\zeta_r \right)\right] \nonumber\\
&\simeq
\left[1-4k(2\alpha_1+\alpha_2)\bar{F}^2 \right]^2
\zeta^i \zeta_i \bar{g}^{tt} \bar{g}^{rr} \left[\zeta^2 - 8k(2\alpha_1+\alpha_2)\bar{F}^2\left(\zeta^t\zeta_t+\zeta^r\zeta_r \right) \right] .
\end{align} 
Then, $\det \mathcal{M}_s=0$ is satisfied when 
\begin{align} 
 \Bigg(\bar{g}^{\mu\nu} - 16 k (2\alpha_1+\alpha_2) \bar{F}^{\mu \rho} \bar{F}^{\nu}{}_{\rho}  \Bigg) \zeta_\mu \zeta_\nu = 0
\end{align}
holds, implying that
the inverse of the effective metric is
\begin{align}
 (\widehat{g}{~}^{-1})^{\mu\nu} = \bar{g}^{\mu\nu} -  8k \left(4\alpha_1+2\alpha_2\right) \bar{F}^{\mu \rho} \bar{F}^{\nu}{}_{\rho} .
\end{align} 

\bibliographystyle{JHEP}  
\bibliography{ref}  

\end{document}